\documentclass[%
reprint,
amsmath,amssymb,
aps,prb,superscriptaddress,floatfix
]{revtex4-2}

\usepackage{graphicx}
\usepackage{amsmath}
\usepackage{amssymb}
\usepackage{dcolumn}
\usepackage{bm}
\usepackage{bbm}
\usepackage{braket}

\usepackage[colorlinks=true,urlcolor=blue,linkcolor=blue,citecolor=blue]{hyperref}

\begin{document}
	
	\title{Formation and Evolution of the Spin-Charge-Entangled Screening Cloud in the Majorana-Kondo System}

	\author{Yun Chen}
	\thanks{Y. Chen and H. Shen contributed equally to this work.}
	\affiliation{National Laboratory of Solid State Microstructures and School of Physics, Nanjing University, Nanjing 210093, China}
	\affiliation{Collaborative Innovation Center of Advanced Microstructures, Nanjing University, Nanjing 210093, China}

	\author{Haojie Shen}
	\thanks{Y. Chen and H. Shen contributed equally to this work.}
	\affiliation{National Laboratory of Solid State Microstructures and School of Physics, Nanjing University, Nanjing 210093, China}
	\affiliation{Collaborative Innovation Center of Advanced Microstructures, Nanjing University, Nanjing 210093, China}
	\affiliation{School of Physics and Astronomy, Shanghai Jiao Tong University, Shanghai 200240, China}


	\author{Wei Su}
	\email{suwei@sicnu.edu.cn}
	\affiliation{College of Physics and Electronic Engineering, Center for Computational Sciences, Sichuan Normal University, Chengdu 610068, China}
	
	\begin{abstract}
		Side-coupled Majorana zero modes in Kondo systems realize a simple yet nontrivial hybridization setup that leads to distinct physics from the conventional Kondo effect. 
		We have demonstrated in a previous work that the system can be described by a spin-charge-entangled (SCE) quantum impurity model with an Andreev$\otimes$normal boundary condition. 
		Here we investigate in detail the formation process and microscopic mechanism of the SCE screening cloud using the numerical renormalization group method. 
		We introduce temperature-dependent spatially integrated correlation functions that provide an unambiguous diagnostic of different components of the SCE screening cloud beyond the impurity entropy and local density of states. 
		Our results reveal a crossover from a sequential two-stage screening of the spin and charge components to a simultaneous screening controlled by a single parameter. 
		We also study the evolution of the low-energy fixed points in the presence of competing terms that break different symmetries and drive the system to different infrared fixed points.
		Our results suggest that the SCE screening effect in the Majorana-Kondo system makes itself a route to detecting Majorana zero modes.
	\end{abstract}
	
	\date{\today}
	\maketitle
	
\section{Introduction}
	Majorana zero modes (MZMs)~\cite{majorana1937,wilczek2009}, characterized by self-conjugation and zero-energy excitation, represent a frontier in condensed matter. Crucially, obeying non-Abelian statistics and protected by nonlocal topology, MZMs are promising candidates for quantum computation~\cite{read2000,kitaev2003,nayak2008}. 
	Experimentally, Majorana signatures have been reported in a variety of setups, such as one-dimensional (1d) superconductor-proximitized Rashba nanowires~\cite{prada2020andreev,mourik2012,das2012zero,rokhinson2012fractional,nadj2014observation,albrecht2016exponential,lutchyn2018}, and 2d topological superconductor (TSC) surfaces~\cite{sun2017detection,wang2012coexistence,williams2012unconventional,xu2014artificial,xu2015experimental, liu2018robust,zhang2018observation,wang2018evidence,chen2019quantized,zhu2020nearly,chen2020robust}. 
	To overcome the fragility of fractional signatures when MZMs are directly contacted to a normal lead (NL) in transport measurements, the quantum dot (QD)-assisted setups have been proposed~\cite{tewari2008testable,leijnse2011scheme,flensberg2011non}. The Kondo resonance~\cite{kondo1964,wilson1975,hewson1993,coleman2015} can be dramatically modified by the MZMs via anti-Fano interference over a broad parameter range, making the MZM-QD-NL platform promising for MZM detection~\cite{tewari2008testable,leijnse2011scheme,flensberg2011non,cheng2014interplay}. 
	
	Although characteristic signatures such as spectral and correlation properties have been obtained using many-body methods such as perturbative renormalization group~\cite{golub2011kondo,cheng2014interplay}, numerical renormalization group (NRG)~\cite{wilson1975,peters2006numerical,weichselbaum2007sumrule,bulla2008} and density matrix renormalization group (DMRG)~\cite{white1992density,white1992real,schollwock2005density,schollwock2011density}, the underlying mechanism that drives the interplay between the Kondo screening and the QD-MZM coupling has long been debated~\cite{golub2011kondo,lee2013kondo,cheng2014interplay,vernek2014subtle,silva2020robustness,majek2022majorana,bollmann2024topological}. On the one hand, at low temperatures, the system is subject to an Andreev$\otimes$normal boundary condition (ANBC), i.e., an Andreev BC for spin-\(\uparrow\) electrons and a normal BC for spin-\(\downarrow\) electrons, due to the QD-MZM coupling~\cite{clark2000andreevscattering,law2009majorana,deacon2010kondo,pillet2013tunneling,fidkowski2012universal,affleck2013topological,cheng2014interplay}. The ANBC leads to the breakdown of the Kondo resonance peak and the violation of the Friedel sum rule for the local density of states (LDoS)~\cite{hewson1993,lee2013kondo,cheng2014interplay}. On the other hand, the entropy of the MZM-QD subsystem (\(S_{\rm imp}\)) suggests that the Kondo screening process is preserved, with the screening regime even extending to zero Coulomb repulsion \(U\)~\cite{cheng2014interplay,silva2020robustness}. Recently, we demonstrated that a spin-charge-entangled (SCE) Kondo effect provides a more appropriate description of the low-energy physics than the conventional spin Kondo effect~\cite{shen2025}. The MZM-QD-NL system can be viewed as a quantum impurity model, with the MZM-QD subsystem playing the role of a composite quantum impurity and the NL acting as a free fermionic bath. In the strong-coupling regime, SCE Kondo singlets form, and the MZM-QD subsystem reduces to an effective Majorana fermion in the infrared (IR) limit, denoted by \(\tilde{\gamma}\)~\cite{shen2025}, which is further coupled to the rest of the NL, resulting in a two-channel-Kondo-like behavior~\cite{emery1992mapping,affleck1995conformal,lopes2020anyons,kattel2024overscreened,tang2025twochannel}. 
	
	However, a full description of the SCE Kondo effect remains incomplete. 
	First, our previous work established the SCE picture of the Majorana-Kondo system, but the formation process of the SCE screening cloud has not been clarified. In particular, while the impurity entropy and the LDoS exhibit a clear multistage behavior, they do not directly resolve how the spin and charge degrees of freedom (DoFs) are screened as the temperature is lowered.
	Second, competing terms must be taken into account. The description of the SCE Kondo effect relies largely on the \({\rm SU}_{\bf L}(2)\) rotational symmetry and the \(\mathbb{Z}_2\) topological degeneracy, whereas sufficiently strong perturbations can destroy the formed \({\rm SU}_{\bf L}(2)\) singlets or modify the ANBC, both of which are commonly encountered in realistic experiments~\cite{prada2012transport,dassarma2012splitting,deng2016coupledqd,deng2018nonlocality}. 

	In this work, we reveal the formation of the SCE screening cloud and its response to symmetry-breaking perturbations by introducing finite-temperature correlation functions that separately track the spin and charge components of the screening cloud. 
	
	\begin{figure}[t]
		\centering
		\includegraphics[width=0.9\linewidth]{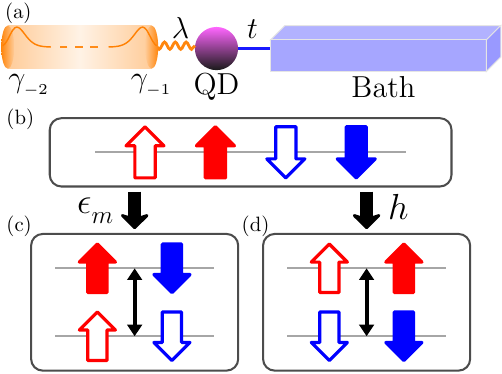}
		\caption{Schematic plots of the system. (a) The 1d setup consists of a normal lead (blue), a quantum dot (violet), and a TSC wire (orange) with two MZMs \(\gamma_{-1}\) and \(\gamma_{-2}\). 
		(b-d) Local energy spectrum of the MZM-QD subsystem. Red and blue colors represent \(z\)-component of the local moment \(L^z=\pm\frac{1}{2}\), and hollow and solid shapes represent the parity \(P_f=0/1\). 
		(b) In the \(\mathbbm{Z}_2\times {\rm SU}_{\bf L}(2)\)-symmetric case, the fourfold degeneracy is labeled by \((P_f,L^z)\)~\cite{shen2025}. 
		(c) A Majorana overlap \(\epsilon_m\) breaks the \(\mathbbm{Z}_2\) topological degeneracy. 
		(d) An effective Zeeman field \(h\) breaks the \({\rm SU}_{\bf L}(2)\) degeneracy. 
		}
		\label{fig:setup}
	\end{figure}

\section{Model and methods}
	The system we consider is shown in Fig.~\ref{fig:setup}(a). The TSC hosts a pair of MZMs \(\gamma_{-1}\), \(\gamma_{-2}\) at the boundaries, with one coupled to the QD. The symmetric Hamiltonian is given by
	\begin{equation}
	H_{0}=\mathrm{i} J_M \gamma_{-1} {\bf L}_{\textrm{QD}}\cdot\bm{\xi}_{0}+J_K{\bf L}_{\textrm{QD}}\cdot{\bf S}_{0} + \sum_{\bf k} \epsilon_{\bf k} c_{{\bf k}\sigma}^\dagger c_{{\bf k}\sigma},
	\label{eq:heff1}
	\end{equation}
	where \(\gamma_{-1} \equiv (f + f^\dagger)\) is the MZM operator coupled to the QD. The QD is described by a local SCE moment \({\bf L}_{\rm QD}\) as emphasized in Ref.~\cite{shen2025}, \(\bm{\xi}_0\) is an irreducible tensor operator of the bath \({\rm SU}_{\bf L}(2)\) group~\cite{shen2025,appendix}, and \(c_{{\bf k}\sigma}^{(\dagger)}\) are bath electron annihilation (creation) operators. The local \({\rm SU}_{\bf L}(2)\) rotational symmetry is generated by \( {\bf L}_i\equiv  {\bf S}_i + \bm{\eta}_i\), with \({\bf S}_i\) ($\bm{\eta}_i$) the generator of the spin (charge) \({\rm SU}(2)\) symmetry at site \(i\in[-2,\infty)\). 
	The local Hamiltonian of the MZM-QD subsystem hosts \(\mathbbm{Z}_2\times {\rm SU}_{\bf L}(2)\) symmetry, thus the energy spectrum displays a 4-fold degeneracy, as demonstrated in Fig.~\ref{fig:setup}(b). 

	Consider symmetry-breaking terms
	\begin{equation}
		H^\prime = \epsilon_m i \gamma_{-1}\gamma_{-2} + h L_{\rm QD}^{z}, 
		\label{eq:heff2}
	\end{equation}
	where \(\gamma_{-2} \equiv (f-f^\dagger)/i\) is the MZM at the other side of the TSC. The MZM coupling \(\epsilon_m\) describes the wavefunction overlap between the two MZMs, which can be induced by finite-size effects. The Zeeman field \(h\) can be induced by a local magnetic field or local particle-hole symmetry breaking~\cite{appendix}. Each term reduces the 4-fold degeneracy to 2-fold, as shown in Fig.~\ref{fig:setup}(c,d). 

	We implement a high-precision full-density-matrix (FDM)-NRG method based on the matrix product state framework to study this composite quantum impurity problem~\cite{weichselbaum2012tensor,merker2012fdm}. 
	We mainly focus on the properties of the strong coupling fixed points and their evolution when physical conditions change. Different quantities are calculated for different purposes. The impurity thermodynamic entropy is defined by the difference between the entropy of the total system and that of only the free bath, \(S_{\rm imp}\equiv S_{\rm tot}-S_{\rm bath}\).
	The normal LDoS is obtained by \(\rho_{\sigma}(\omega) = -\frac{1}{\pi}{\rm Im}\mathcal{T}_{\sigma}(\omega) = -\frac{1}{\pi}{\rm Im}\braket{\braket{O_{\sigma}|O_{\sigma}^\dagger}}_{\omega}\). The scattering operator \(O_{\sigma}\) contains three parts~\cite{appendix},
	\begin{gather}
			K_{\sigma} =  J_K (\sigma L^z c_{0\sigma} + L_{\bar{\sigma}} L_{0\bar{\sigma}}), \\
			Q_{\sigma} = J_Q (c_{0\sigma}L^z + \sigma c_{0\bar{\sigma}}^\dagger L_+), \\
			M_{\uparrow} = M_{\uparrow}^\dagger  = J_M L^z \gamma,\quad M_{\downarrow} = J_M L_+ \gamma,
	\end{gather}
	where \(K_{\sigma}\) is exactly the Kondo scattering operator~\cite{costi2000kondo}.
	Apart from the normal LDoS, there is an anomalous component \(A_{\uparrow}=-\frac{1}{\pi}{\rm Im}\braket{\braket{O_{\sigma}|O_{\sigma}}}_{\omega}\), corresponding to the off-diagonal term of the \(\mathcal{T}\)-matrix~\cite{lee2013kondo,shen2025,appendix}. In the presence of \({\rm SU}_{\bf L}(2)\) symmetry, one has \(\rho_\uparrow + A_\uparrow = \rho_\downarrow\)~\cite{shen2025}.

	To uncover the temperature-dependent behavior of the SCE screening, we develop our FDM-NRG method and introduce the spatially integrated correlation functions between the SCE impurity and the bath, \(I_{LL}(T) = I_{LS}(T) + I_{L\eta}(T)\).
	The exact definitions are
	\begin{equation}
		\begin{aligned}
		I_{LS}(T) &= \int_{\rm bath} dx ( \langle {\bf L}_{\rm QD}\cdot {\bf S}(x) \rangle - \langle {\bf L}_{\rm QD} \rangle \cdot \langle {\bf S}(x) \rangle ),\\
		I_{L\eta}(T) &= \int_{\rm bath} dx ( \langle {\bf L}_{\rm QD}\cdot {\bm \eta}(x) \rangle - \langle {\bf L}_{\rm QD} \rangle \cdot \langle {\bm \eta}(x) \rangle ),
		\end{aligned}
	\end{equation}
	where the second terms vanish if \(h=0\).
	They quantify the screening of the SCE impurity \({\bf L}_{\rm QD}\) by the spin and charge DoFs of the bath, respectively.
	In practical calculations, the integral over \(x\) is replaced by a summation over the bath sites \(i\) of the Wilson chain, which can be obtained by a mapping of the original Hamiltonian~\cite{wilson1975,hewson1993,bulla2008}. 
	We point out that, though zero-temperature spin-spin correlator \(\braket{{\bf S}(0) \cdot {\bf S}(x)}\) in the Kondo system has been well studied with DMRG to identify the correlation length of the Kondo cloud~\cite{nishimoto2004density,hand2006spinrelations,holzner2009kondo}, our \(T\)-dependent correlators unambiguously capture different strong-coupling characteristic temperature scales when several competing operators coexist, and directly resolve how different components of the screening cloud build up as the temperature is lowered. This method can be applied to a broad class of composite or artificial quantum impurity systems~\cite{jones1988twoimpurity,zitko2006enhanced,craig2004tunable,sasaki2006nonlocal,bork2011twoimpurity}.

\section{SCE screening and characteristic temperatures}
	\begin{figure}[t]
		\centering
		\includegraphics[width=\linewidth]{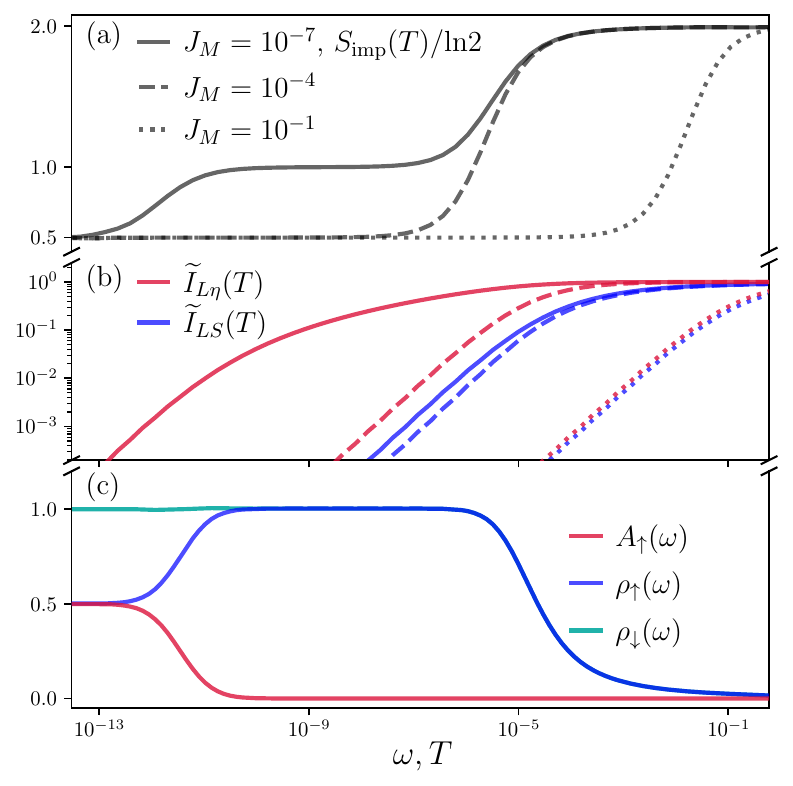}
		\caption{Results for the \(\mathbbm{Z}_2\times {\rm SU}_{\bf L}(2)\) symmetric case for \(J_K=0.2\) and \(J_M=10^{-1},10^{-4},10^{-7}\).
		(a) The impurity entropy \(S_{\rm imp}(T)\) demonstrates the multistage behavior of the composite quantum impurity model. 
		(b) The reduced \(T\)-dependent correlation functions \(\widetilde{I}_{LS/L\eta}(T)\) define two different characteristic temperatures, \(T_{\rm K}\) and \(T_{\rm SCE}\), and reveal a crossover from sequential screening to simultaneous screening.
		(c) The anomalous and normal LDoSs \(A_{\sigma}/\rho_{\sigma}(\omega)\) indicate an ANBC in the IR limit (plotted only for \(J_M=10^{-7}\) for brevity).
		}
		\label{fig:jm}
	\end{figure}
	We first briefly summarize the SCE screening effect in the \(\mathbbm{Z}_2\times {\rm SU}_{\bf L}(2)\) symmetric case. The low-energy properties are well understood in terms of SCE singlets subject to the ANBC~\cite{shen2025}.
	In Fig.~\ref{fig:jm}(a-c) we plot the impurity thermodynamic entropy \(S_{\rm imp}(T)\), the reduced correlation functions \(\widetilde{I}_{LS/L\eta}(T)=1-\frac{I_{LS/L\eta}(T)}{I_{LS/L\eta}(0)}\), and the LDoS \(A_{\sigma}/\rho_{\sigma}(\omega)\) for three values of the MZM-QD coupling \(J_M\). The entropy and the LDoS were also calculated in Ref.~\cite{shen2025}, where the two characteristic temperatures \(T_{\rm K}\) and \(T_{\rm A \otimes N}\) were defined, respectively, by the disappearance of the local-moment fixed point (\(S_{\rm imp}(T)=2\ln2 \to \ln2\)) and by the appearance of the anomalous LDoS (\(A_{\sigma}(\omega)=\frac{1}{2}\))~\cite{lee2013kondo,shen2025}. 
	For a small \(J_M=10^{-7}\) (solid lines in Fig.~\ref{fig:jm}), there are two strong-coupling fixed points in addition to the local-moment fixed point (\(S_{\rm imp}=2\ln2\)). A plausible explanation for this multistage behavior is that the QD and the MZM \(\gamma_{-1}\) are screened sequentially by the NL bath. In this picture, the first strong-coupling fixed point with \(S_{\rm imp}=\ln 2\) resembles a conventional Kondo fixed point, where the SCE local moment \({\bf L}_{\rm QD}\) is screened. 
	As the temperature is further lowered, the MZM-induced ANBC sets in and the Kondo fixed point becomes unstable. The MZM \(\gamma_{-1}\) is then screened and the energy scale \(T_{\rm A\otimes N}\sim 10^{-12}\) appears as the signature of the second strong-coupling fixed point, whereas \(\gamma_{-2}\) remains decoupled and contributes an entropy \(S_{\rm imp}=\frac{1}{2}\ln 2\). 

	Nevertheless, the explanation above is unreliable beacuse (i) the impurity entropy \(S_{\rm imp}\) defined by subtraction only counts the active boundary DoFs and does not explain where these DoFs come from. (ii) The anomalous LDoS \(A_{\sigma}(\omega)\) identifies the existence of the ANBC, but it is not a direct probe of the screening process either. 
	Therefore, these quantities do not resolve the microscopic distinction between different strong-coupling fixed points. In the following, we study these regimes in detail by calculating the \(T\)-dependent correlation functions, which are intrinsic to the screening physics. They also provide more appropriate definitions of the characteristic temperature scales and more precise descriptions of the screening process.

	Results for the correlation functions \(\widetilde{I}_{LS/L\eta}(T)\) are plotted in Fig.~\ref{fig:jm}(b). 
	For small \(J_M=10^{-7}\) (solid lines), the two correlation functions \(\widetilde{I}_{L\eta}(T)\) and \(\widetilde{I}_{LS}(T)\) are clearly separated, demonstrating that the charge and spin components of the SCE screening cloud are established asynchronously. 
	The spin screening develops first at \(T\sim 10^{-7}\), where the reduced \(L\)-\(S\) correlation \(\widetilde{I}_{LS}(T)\) decays as a power law and approaches zero, namely, the correlation function \(I_{LS}(T)\) saturates  to \(I_{LS}(T=0)\). The charge screening builds up at a lower scale \(T\sim 10^{-12}\), where \(\widetilde{I}_{L\eta}(T)\) approaches zero and the SCE local moment \({\bf L}_{\rm QD}\) becomes fully screened. In this sense, the two strong-coupling temperatures are defined as \(T_{\rm K} \propto D e^{-1/(\rho_0 J_K)} \sim 10^{-7}\), the spin Kondo screening temperature, and \(T_{\rm SCE} \propto \rho_0 J_M^2 \sim 10^{-12}\), the SCE screening temperature. 
	For intermediate \(J_M=10^{-4}\), the two correlation functions still split (dashed lines), yet the \(S_{\rm imp}=\frac{1}{2}\ln 2\) plateau does not appear and we have \(T_{\rm SCE}\lesssim T_{\rm K}\).
	In contrast, for large \(J_M=10^{-1}\) (dotted lines), the two correlation functions nearly overlap, indicating simultaneous screening of the charge and spin DoFs. The MZM-QD coupling \(J_M\) dominates over the Kondo coupling \(J_K\), so the SCE energy scale satisfies \(T_{\rm SCE} \gg T_{\rm K}\), and the spin Kondo temperature \(T_{\rm K}\) is no longer identifiable. 

\section{MZM overlap and SCE screening without ANBC}
	\begin{figure}[t]
		\centering
		\includegraphics[width=\linewidth]{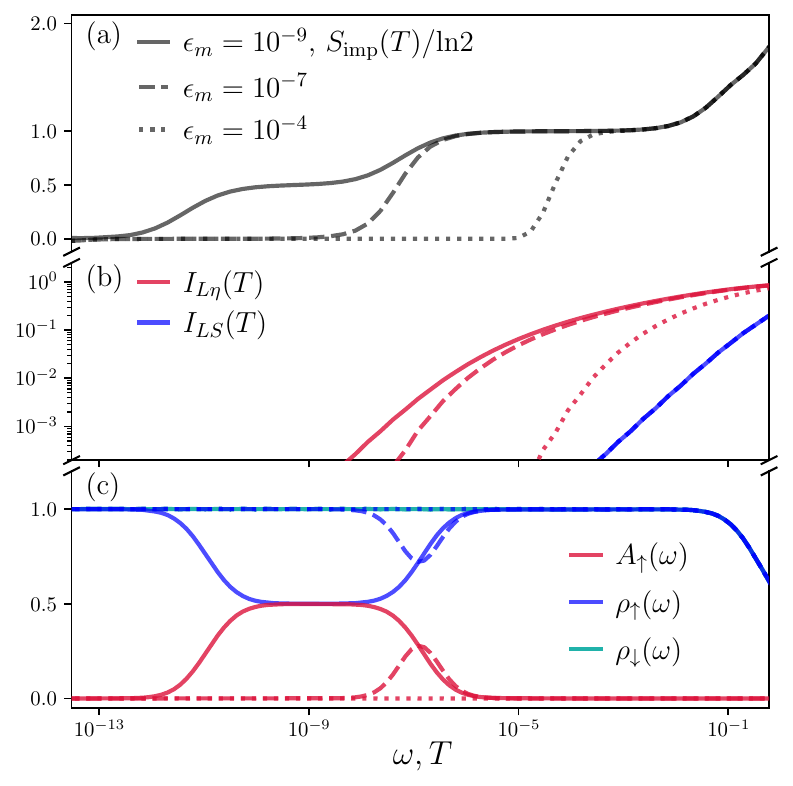}
		\caption{Results for the \(\mathbb{Z}_2\)-symmetry breaking case. The physical parameters are \(J_K=1\), \(J_M=10^{-4}\), and \(\epsilon_m=10^{-9}, 10^{-7}, 10^{-4}\).
		(a) The impurity entropy \(S_{\rm imp}(T)\) decreases to 0 at zero temperature. A large overlap merges the pair of MZMs into a spinless fermion and removes the \(S_{\rm imp}=1/2\ln2\) plateau. 
		(b) The reduced \(T\)-dependent correlation functions \(\widetilde{I}_{LS/L\eta}(T)\). The evolution of \(\widetilde{I}_{L\eta}(T)\) identifies the SCE screening process even when the ANBC is absent.
		(c) The anomalous and normal LDoSs \(A_{\sigma}/\rho_{\sigma}(\omega)\). 
		}
		\label{fig:em}
	\end{figure}
	The finite-size effect of the TSC wire results in an overlap between the two MZMs at the boundaries, i.e., \(i\epsilon_m \gamma_{-2} \gamma_{-1}\). The coupling strength \(\epsilon_m\) can be quantified by the length \(L\) of the TSC wire, \(\epsilon_m\sim \int \psi_{-1}(x)H_{\rm TSC}\psi_{-2}(x) {\rm d}x\propto e^{-L/\xi}\), where \(\xi\) is the superconducting coherence length. 
	
	In Fig.~\ref{fig:em}(a-c), we calculate the impurity entropy, the correlation functions, and the LDoS for varying \(\epsilon_m\). At low temperatures, the impurity entropy \(S_{\rm imp}(T)\) decreases from \(\frac{1}{2}\ln2\) to 0, introducing an additional energy scale \(T_{\rm M}\) (\(\sim 10^{-12}\) for \(\epsilon_m=10^{-9}\)). Meanwhile, the normal component of the LDoS \(\rho_{\uparrow}\) returns to 1 from \(\frac{1}{2}\), and the anomalous LDoS \(A_{\uparrow}\) correspondingly falls back to 0 from \(\frac{1}{2}\), indicating that the MZM-MZM coupling lifts the \(\mathbb{Z}_2\) degeneracy (Fig.~\ref{fig:setup}(c)) and that the ANBC breaks down when \(T<T_{\rm M}\). The system therefore exhibits a three-stage behavior, as shown in Fig.~\ref{fig:em}(a,c). 
	As \(\epsilon_m\) increases, the energy scale \(T_{\rm M}\) rises above \(T_{\rm SCE}\), and the \(S_{\rm imp}(T)=\frac{1}{2}{\rm ln2}\) plateau disappears. For \(\epsilon_m=10^{-7}\) and \(10^{-4}\) (see the dashed and dotted curves in Fig.~\ref{fig:em}(a)), the two MZMs bind to form a spinless fermion and are absorbed into the NL bath, driving the boundary entropy \(S_{\rm imp}(T)\) to decrease directly from \(\ln2\) to 0. 
	
	The SCE screening effect, however, persists down to low temperatures \(T < T_{\rm M}\) even after the disappearance of the \(S_{\rm imp}={\rm ln2}\) plateau and the \(A_{\uparrow}(\omega)=\frac{1}{2}\) plateau, as shown in Fig.~\ref{fig:em}(b) (red curves). This demonstrates that the existence of the SCE screening cloud does not rely on the presence of the ANBC itself. 
	For small \(\epsilon_m=10^{-9}\), where \(T_{\rm M} \ll T_{\rm SCE}\), the correlation function \(I_{L\eta}(T)\) saturates at \(T_{\rm SCE}\) rather than at \(T_{\rm M}\), indicating SCE screening with an ANBC. For large \(\epsilon_m\), such that the distinct scale \(T_{\rm SCE}\) disappears, \(I_{L\eta}(T)\) saturates at \(T_{\rm M}\) without an ANBC. In other words, the SCE screening effect is robust against the parity-symmetry-breaking perturbation. 
	It is also worth noting that for large \(\epsilon_m=10^{-4}\) (the dotted curves), the anomalous LDoS vanishes, \(A_{\uparrow}(\omega) \equiv 0\), and the \(T\)-dependent correlation functions \(I_{LS}(T)\) and \(I_{L\eta}(T)\) become the only quantities that distinguish between the spin screened fixed point and the SCE screened fixed point.

	\begin{figure}[htbp]
		\centering
		\includegraphics[width=\linewidth]{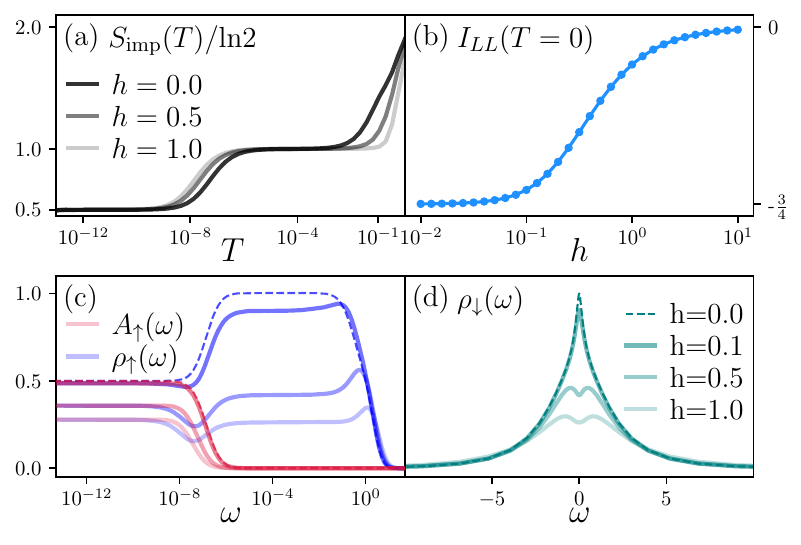}
		\caption{Results for the \({\rm SU}_{\bf L}(2)\)-symmetry breaking cases for \(J_K=1\), \(J_M=10^{-4}\), and \(h=0.0,0.5,1.0\) for (a), \(h \in [10^{-2}, 10^{1}]\) for (b), and \(h=0.0,0.1,0.5,1.0\) for (c-d).
		(a) The impurity entropy \(S_{\rm imp}\) for \(h=0,0.5,1.0\). For large \(h\), the \(S_{\rm imp}=\ln 2\) plateau refers to polarized states instead of singlets. 
		(b) The evolution of the zero-temperature correlation function \(I_{LL}(T=0)\) indicates the disappearance of the \({\bf L}\)-singlet as \(h\) increases.
		(c-d) The splitting of the LDoSs \(A_{\uparrow}(\omega)\), \(\rho_{\uparrow}(\omega)\) and \(\rho_{\downarrow}(\omega)\) for \(h=0,0.1,0.5,1.0\). 
		}
		\label{fig:gb}
	\end{figure}
	
\section{Pseudospin polarization and the SCE breakdown}
	Now we turn to the pseudospin polarization term \(h L^z\). This term can be introduced by an external Zeeman field, or emerges from the particle-hole asymmetry of the QD~\cite{appendix}. This term breaks the \({\rm SU}_{\bf L}(2)\) symmetry, but preserves the topological degeneracy, which can be directly seen from the \(S_{\rm imp}(T)=\frac{1}{2}{\rm ln2}\) plateau in Fig.~\ref{fig:gb}(a) even for a large \(h\). The spectrum of the MZM-QD subsystem is demonstrated in Fig.~\ref{fig:setup}(d). 

	From the perspective of the total LDoS \(\rho(\omega)=A_{\uparrow}+\rho_{\uparrow}+\rho_{\downarrow}\), the influence of the Zeeman field on the SCE Kondo physics is akin to that on the conventional Kondo resonance peak~\cite{costi2000kondo,hewson1993,bulla2008}. 
	When \(h < T_{\rm K}\), one can expect the SCE \({\rm SU}_{\bf L}(2)\) singlets to survive to some extent, and the SCE Kondo resonance peak remains unsplit, although its height decreases slightly, as shown in Fig.~\ref{fig:gb}(d) for \(h=0.1\). Upon increasing \(h\), the LDoS \(\rho(\omega)\) splits and a dip appears at \(\omega=0\), indicating that the many-body singlet is destroyed. This is consistent with the zero-temperature behavior of the total screening cloud \(I_{LL}(T=0)=I_{LS}(T=0)+I_{L\eta}(T=0)\), as shown in Fig.~\ref{fig:gb}(b). The saturated value of \(I_{LL}(T=0)\) decreases from \(-\frac{3}{4}\) (singlet) to 0 (polarized) as the Zeeman field \(h\) increases, especially when \(h > T_{\rm K} \sim 10^{-1}\). 

	In Fig.~\ref{fig:gb}(a), the behavior of the impurity entropy \(S_{\rm imp}(T)\) appears nearly unchanged. However, the plateau of \(S_{\rm imp} = \ln 2\) has different physical meanings for different values of \(h\). It represents a SCE screened strong coupling fixed point for \(h\ll T_{\rm K}\), but a polarized single-body fixed point for \(h\gg T_{\rm K}\). In both cases, the far-end MZM \(\gamma_{-2}\) is decoupled from the \(\gamma_{-1}\)-QD-NL subsystem, so the energy scale of \(T_{\rm SCE}\) changes negligibly even for large \(h\). The robustness of the ANBC can be easily understood in both limits when \(h\) is very small and very large. 
	For \(h\ll T_{\rm K}\), the ground states of the MZM-QD-\(c_0\) subsystem are two degenerate \({\rm SU}_{\bf L}(2)\) SCE singlets with different parities \(P_f\). 
	When \(h\gg T_{\rm K}\) and \(h\) becomes the most relevant operator, the ground states are two polarized states, and are also degenerate with opposite parities. 
	This is expected since the Zeeman term commutes with the parity operator, \([L^z, P_f] \equiv 0\), so that the ground states always preserve the topological degeneracy and are not affected by the \(L^z\) polarization~\cite{appendix}.

\section{Conclusions and discussions}
	In this work, we study the screening physics of a composite quantum impurity system with multiple competing ingredients, including a Kondo interaction, a Majorana-Kondo coupling, a MZM-MZM overlap and a Zeeman field. 
	By combining analytical analysis with numerical calculations, our work complements the SCE Kondo description of this quantum impurity problem. The formation process of the SCE screening cloud is uncovered by computing the temperature-dependent correlation functions \(I_{LS/L\eta}(T)\). In different regimes, the spin and charge DoFs can be screened either asynchronously or simultaneously. 
	We also clarify how different symmetry-breaking terms alter the SCE screening process.
	The non-Fermi liquid fixed point is protected jointly by the ANBC and the \(\mathbb{Z}_2\) symmetry but independent of the \({\rm SU}_{\bf L}(2)\) symmetry, and the SCE singlet is protected by the \({\rm SU}_{\bf L}(2)\) symmetry. 
	Our analysis based on correlation functions offers an effective route to understanding the screening physics of generic composite impurity systems. 
	Taking the superconducting bulk into account may lead to additional phenomena, such as YSR-like states~\cite{luh1965,shiba1968classical,rusinov1969superconductivity,pasnoori2020kondo,pasnoori2022rise,moca2025fractionalized,moca2021kondo,huang2025largeN,kattel2026thermodynamics}.

	In a recent work, the charge Kondo screening cloud was experimentally observed through oscillations of the QD conductance, from which the universal scaling of the screening correlation length $\xi_{\rm K}$ and the Kondo temperature $T_{\rm K}$ was extracted~\cite{ivan2020observation,affleck2010kondo, park2013directly}. 
	In our MZM-QD-NL setup, the presence of MZMs gives rise to an SCE screening mechanism and additional characteristic energy scales $T_{\rm SCE}$ and $T_{\rm M}$, which are expected to produce scaling behaviors distinct from that of the conventional Kondo effect. 
	Our work therefore suggests a feasible experimental route to detecting MZMs.

\begin{acknowledgments}
	We thank Rui Wang and Tigran Sedrakyan for valuable discussions. This work is supported by the National Natural Science Foundation of China (Grants No. 125B2076), the National Natural Science Foundation of China (No. 11904245) and the Postdoctoral Science Foundation of China (No. 2021M690330). 
\end{acknowledgments}

	\bibliographystyle{apsrev4-2}
	\bibliography{KondoMajorana}

@article{majorana1937,
	title={Teoria simmetrica dell’elettrone e del positrone},
	author={Majorana, Ettore},
	journal={Il Nuovo Cimento},
	volume={14},
	number={4},
	pages={171--184},
	year={1937},
	doi={10.1007/BF02961314},
	url={https://doi.org/10.1007/BF02961314},
	publisher={Springer}
}

@article{wilczek2009,
	title={Majorana returns},
	author={Wilczek, Frank},
	journal={Nat. Phys.},
	volume={5},
	number={9},
	pages={614--618},
	year={2009},
	doi={10.1038/nphys1380},
	url={https://doi.org/10.1038/nphys1380},
	publisher={Nature Publishing Group UK London}
}

@article{read2000,
	title={Paired states of fermions in two dimensions with breaking of parity and time-reversal symmetries and the fractional quantum Hall effect},
	author={Read, Nicholas and Green, Dmitry},
	journal={Phys. Rev. B},
	volume={61},
	number={15},
	pages={10267--10297},
	year={2000},
	doi={10.1103/PhysRevB.61.10267},
	url={https://doi.org/10.1103/PhysRevB.61.10267},
	publisher={APS}
}

@article{kitaev2003,
	title={Fault-tolerant quantum computation by anyons},
	author={Kitaev, A Yu},
	journal={Ann. Phys.},
	volume={303},
	number={1},
	pages={2--30},
	year={2003},
	doi={10.1016/S0003-4916(02)00018-0},
	url={https://doi.org/10.1016/S0003-4916(02)00018-0},
	publisher={Elsevier}
}

@article{nayak2008,
	title={Non-Abelian anyons and topological quantum computation},
	author={Nayak, Chetan and Simon, Steven H and Stern, Ady and Freedman, Michael and Das Sarma, Sankar},
	journal={Rev. Mod. Phys.},
	volume={80},
	number={3},
	pages={1083--1159},
	year={2008},
	doi={10.1103/RevModPhys.80.1083},
	url={https://doi.org/10.1103/RevModPhys.80.1083},
	publisher={APS}
}

@article{prada2020andreev,
	title={From Andreev to Majorana bound states in hybrid superconductor--semiconductor nanowires},
	author={Prada, Elsa and San-Jose, Pablo and de Moor, Michiel WA and Geresdi, Attila and Lee, Eduardo JH and Klinovaja, Jelena and Loss, Daniel and Nyg{\aa}rd, Jesper and Aguado, Ram{\'o}n and Kouwenhoven, Leo P},
	journal={Nat. Rev. Phys.},
	volume={2},
	number={10},
	pages={575--594},
	year={2020},
	doi={10.1038/s42254-020-0228-y},
	url={https://doi.org/10.1038/s42254-020-0228-y},
	publisher={Nature Publishing Group UK London}
}

@article{mourik2012,
	title={Signatures of Majorana fermions in hybrid superconductor-semiconductor nanowire devices},
	author={Mourik, Vincent and Zuo, Kun and Frolov, Sergey M and Plissard, SR and Bakkers, Erik PAM and Kouwenhoven, Leo P},
	journal={Science},
	volume={336},
	number={6084},
	pages={1003--1007},
	year={2012},
	doi={10.1126/science.1222360},
	url={https://doi.org/10.1126/science.1222360},
	publisher={American Association for the Advancement of Science}
}

@article{das2012zero,
	title={Zero-bias peaks and splitting in an Al--InAs nanowire topological superconductor as a signature of Majorana fermions},
	author={Das, Anindya and Ronen, Yuval and Most, Yonatan and Oreg, Yuval and Heiblum, Moty and Shtrikman, Hadas},
	journal={Nat. Phys.},
	volume={8},
	number={12},
	pages={887--895},
	year={2012},
	doi={10.1038/nphys2479},
	url={https://doi.org/10.1038/nphys2479},
	publisher={Nature Publishing Group UK London}
}

@article{rokhinson2012fractional,
	title={The fractional ac Josephson effect in a semiconductor--superconductor nanowire as a signature of Majorana particles},
	author={Rokhinson, Leonid P and Liu, Xinyu and Furdyna, Jacek K},
	journal={Nat. Phys.},
	volume={8},
	number={11},
	pages={795--799},
	year={2012},
	doi={10.1038/nphys2429},
	url={https://doi.org/10.1038/nphys2429},
	publisher={Nature Publishing Group UK London}
}

@article{nadj2014observation,
	title={Observation of Majorana fermions in ferromagnetic atomic chains on a superconductor},
	author={Nadj-Perge, Stevan and Drozdov, Ilya K and Li, Jian and Chen, Hua and Jeon, Sangjun and Seo, Jungpil and MacDonald, Allan H and Bernevig, B Andrei and Yazdani, Ali},
	journal={Science},
	volume={346},
	number={6209},
	pages={602--607},
	year={2014},
	doi={10.1126/science.1259327},
	url={https://doi.org/10.1126/science.1259327},
	publisher={American Association for the Advancement of Science}
}

@article{albrecht2016exponential,
	title={Exponential protection of zero modes in Majorana islands},
	author={Albrecht, Sven Marian and Higginbotham, Andrew P and Madsen, Morten and Kuemmeth, Ferdinand and Jespersen, Thomas Sand and Nyg{\aa}rd, Jesper and Krogstrup, Peter and Marcus, CM},
	journal={Nature},
	volume={531},
	number={7593},
	pages={206--209},
	year={2016},
	doi={10.1038/nature17162},
	url={https://doi.org/10.1038/nature17162},
	publisher={Nature Publishing Group UK London}
}

@article{lutchyn2018,
	title={Majorana zero modes in superconductor--semiconductor heterostructures},
	author={Lutchyn, Roman M and Bakkers, Erik PAM and Kouwenhoven, Leo P and Krogstrup, Peter and Marcus, Charles M and Oreg, Yuval},
	journal={Nat. Rev. Mater.},
	volume={3},
	number={5},
	pages={52--68},
	year={2018},
	doi={10.1038/s41578-018-0003-1},
	url={https://doi.org/10.1038/s41578-018-0003-1},
	publisher={Nature Publishing Group UK London}
}

@article{sun2017detection,
	title={Detection of Majorana zero mode in the vortex},
	author={Sun, Hao-Hua and Jia, Jin-Feng},
	journal={npj Quantum Mater.},
	volume={2},
	number={1},
	pages={34},
	year={2017},
	doi={10.1038/s41535-017-0037-4},
	url={https://doi.org/10.1038/s41535-017-0037-4},
	publisher={Nature Publishing Group UK London}
}

@article{wang2012coexistence,
	title={The coexistence of superconductivity and topological order in the Bi2Se3 thin films},
	author={Wang, Mei-Xiao and Liu, Canhua and Xu, Jin-Peng and Yang, Fang and Miao, Lin and Yao, Meng-Yu and Gao, CL and Shen, Chenyi and Ma, Xucun and Chen, X and others},
	journal={Science},
	volume={336},
	number={6077},
	pages={52--55},
	year={2012},
	doi={10.1126/science.1216466},
	url={https://doi.org/10.1126/science.1216466},
	publisher={American Association for the Advancement of Science}
}

@article{williams2012unconventional,
	title={Unconventional Josephson effect in hybrid superconductor-topological insulator devices},
	author={Williams, JR and Bestwick, AJ and Gallagher, P and Hong, Seung Sae and Cui, Y and Bleich, Andrew S and Analytis, JG and Fisher, IR and Goldhaber-Gordon, D},
	journal={Phys. Rev. Lett.},
	volume={109},
	number={5},
	pages={056803},
	year={2012},
	doi={10.1103/PhysRevLett.109.056803},
	url={https://doi.org/10.1103/PhysRevLett.109.056803},
	publisher={APS}
}

@article{xu2014artificial,
	title={Artificial topological superconductor by the proximity effect},
	author={Xu, Jin-Peng and Liu, Canhua and Wang, Mei-Xiao and Ge, Jianfeng and Liu, Zhi-Long and Yang, Xiaojun and Chen, Yan and Liu, Ying and Xu, Zhu-An and Gao, Chun-Lei and others},
	journal={Phys. Rev. Lett.},
	volume={112},
	number={21},
	pages={217001},
	year={2014},
	doi={10.1103/PhysRevLett.112.217001},
	url={https://doi.org/10.1103/PhysRevLett.112.217001},
	publisher={APS}
}

@article{xu2015experimental,
	title={Experimental detection of a Majorana mode in the core of a magnetic vortex inside a topological insulator-superconductor Bi 2 Te 3/NbSe 2 heterostructure},
	author={Xu, Jin-Peng and Wang, Mei-Xiao and Liu, Zhi Long and Ge, Jian-Feng and Yang, Xiaojun and Liu, Canhua and Xu, Zhu An and Guan, Dandan and Gao, Chun Lei and Qian, Dong and others},
	journal={Phys. Rev. Lett.},
	volume={114},
	number={1},
	pages={017001},
	year={2015},
	doi={10.1103/PhysRevLett.114.017001},
	url={https://doi.org/10.1103/PhysRevLett.114.017001},
	publisher={APS}
}

@article{liu2018robust,
	title={Robust and clean Majorana zero mode in the vortex core of high-temperature superconductor (Li 0.84 Fe 0.16) OHFeSe},
	author={Liu, Qin and Chen, Chen and Zhang, Tong and Peng, Rui and Yan, Ya-Jun and Wen, Chen-Hao-Ping and Lou, Xia and Huang, Yu-Long and Tian, Jin-Peng and Dong, Xiao-Li and others},
	journal={Phys. Rev. X},
	volume={8},
	number={4},
	pages={041056},
	year={2018},
	doi={10.1103/PhysRevX.8.041056},
	url={https://doi.org/10.1103/PhysRevX.8.041056},
	publisher={APS}
}

@article{zhang2018observation,
	title={Observation of topological superconductivity on the surface of an iron-based superconductor},
	author={Zhang, Peng and Yaji, Koichiro and Hashimoto, Takahiro and Ota, Yuichi and Kondo, Takeshi and Okazaki, Kozo and Wang, Zhijun and Wen, Jinsheng and Gu, Genda D and Ding, Hong and others},
	journal={Science},
	volume={360},
	number={6385},
	pages={182--186},
	year={2018},
	doi={10.1126/science.aan4596},
	url={https://doi.org/10.1126/science.aan4596},
	publisher={American Association for the Advancement of Science}
}

@article{wang2018evidence,
	title={Evidence for Majorana bound states in an iron-based superconductor},
	author={Wang, Dongfei and Kong, Lingyuan and Fan, Peng and Chen, Hui and Zhu, Shiyu and Liu, Wenyao and Cao, Lu and Sun, Yujie and Du, Shixuan and Schneeloch, John and others},
	journal={Science},
	volume={362},
	number={6412},
	pages={333--335},
	year={2018},
	doi={10.1126/science.aao1797},
	url={https://doi.org/10.1126/science.aao1797},
	publisher={American Association for the Advancement of Science}
}

@article{chen2019quantized,
	title={Quantized conductance of Majorana zero mode in the vortex of the topological superconductor (Li0. 84Fe0. 16) OHFeSe},
	author={Chen, C and Liu, Q and Zhang, TZ and Li, D and Shen, PP and Dong, XL and Zhao, Z-X and Zhang, T and Feng, DL},
	journal={Chin. Phys. Lett.},
	volume={36},
	number={5},
	pages={057403},
	year={2019},
	doi={10.1088/0256-307X/36/5/057403},
	url={https://doi.org/10.1088/0256-307X/36/5/057403},
	publisher={IOP Publishing}
}

@article{zhu2020nearly,
	title={Nearly quantized conductance plateau of vortex zero mode in an iron-based superconductor},
	author={Zhu, Shiyu and Kong, Lingyuan and Cao, Lu and Chen, Hui and Papaj, Micha{\l} and Du, Shixuan and Xing, Yuqing and Liu, Wenyao and Wang, Dongfei and Shen, Chengmin and others},
	journal={Science},
	volume={367},
	number={6474},
	pages={189--192},
	year={2020},
	doi={10.1126/science.aax0274},
	url={https://doi.org/10.1126/science.aax0274},
	publisher={American Association for the Advancement of Science}
}

@article{chen2020robust,
	title={Robust zero energy modes on superconducting bismuth islands deposited on Fe (Te, Se)},
	author={Chen, Xiaoyu and Chen, Mingyang and Duan, Wen and Yang, Huan and Wen, Hai-Hu},
	journal={Nano Letters},
	volume={20},
	number={5},
	pages={2965--2972},
	year={2020},
	doi={10.1021/acs.nanolett.9b04639},
	url={https://doi.org/10.1021/acs.nanolett.9b04639},
	publisher={ACS Publications}
}

@article{tewari2008testable,
	title={Testable Signatures of Quantum Nonlocality in a Two-Dimensional Chiral p-Wave Superconductor},
	author={Tewari, Sumanta and Zhang, Chuanwei and Das Sarma, S and Nayak, Chetan and Lee, Dung-Hai},
	journal={Phys. Rev. Lett.},
	volume={100},
	number={2},
	pages={027001},
	year={2008},
	doi={10.1103/PhysRevLett.100.027001},
	url={https://doi.org/10.1103/PhysRevLett.100.027001},
	publisher={APS}
}

@article{leijnse2011scheme,
	title={Scheme to measure Majorana fermion lifetimes using a quantum dot},
	author={Leijnse, Martin and Flensberg, Karsten},
	journal={Phys. Rev. B},
	volume={84},
	number={14},
	pages={140501},
	year={2011},
	doi={10.1103/PhysRevB.84.140501},
	url={https://doi.org/10.1103/PhysRevB.84.140501},
	publisher={APS}
}

@article{flensberg2011non,
	title={Non-Abelian operations on Majorana fermions via single-charge control},
	author={Flensberg, Karsten},
	journal={Phys. Rev. Lett.},
	volume={106},
	number={9},
	pages={090503},
	year={2011},
	doi={10.1103/PhysRevLett.106.090503},
	url={https://doi.org/10.1103/PhysRevLett.106.090503},
	publisher={APS}
}

@article{kondo1964,
	title={Resistance minimum in dilute magnetic alloys},
	author={Kondo, Jun},
	journal={Prog. Theor. Phys.},
	volume={32},
	number={1},
	pages={37--49},
	year={1964},
	doi={10.1143/PTP.32.37},
	url={https://doi.org/10.1143/PTP.32.37},
	publisher={Oxford University Press}
}

@article{wilson1975,
	title={The renormalization group: Critical phenomena and the Kondo problem},
	author={Wilson, Kenneth G},
	journal={Rev. Mod. Phys.},
	volume={47},
	number={4},
	pages={773--840},
	year={1975},
	doi={10.1103/RevModPhys.47.773},
	url={https://doi.org/10.1103/RevModPhys.47.773},
	publisher={APS}
}

@book{hewson1993,
	place={New York},
	title={The Kondo Problem to Heavy Fermions},
	publisher={Cambridge University Press},
	author={Hewson, Alexander Cyril},
	year={1993},
	doi={10.1017/CBO9780511470752},
	url={https://doi.org/10.1017/CBO9780511470752}
}

@article{peters2006numerical,
  title = {Numerical renormalization group approach to Green's functions for quantum impurity models},
  author = {Peters, Robert and Pruschke, Thomas and Anders, Frithjof B.},
  journal = {Phys. Rev. B},
  volume = {74},
  issue = {24},
  pages = {245114},
  numpages = {12},
  year = {2006},
  month = {Dec},
  publisher = {American Physical Society},
  doi = {10.1103/PhysRevB.74.245114},
  url = {https://link.aps.org/doi/10.1103/PhysRevB.74.245114}
}

@article{weichselbaum2007sumrule,
  title = {Sum-Rule Conserving Spectral Functions from the Numerical Renormalization Group},
  author = {Weichselbaum, Andreas and von Delft, Jan},
  journal = {Phys. Rev. Lett.},
  volume = {99},
  issue = {7},
  pages = {076402},
  numpages = {4},
  year = {2007},
  month = {Aug},
  publisher = {American Physical Society},
  doi = {10.1103/PhysRevLett.99.076402},
  url = {https://link.aps.org/doi/10.1103/PhysRevLett.99.076402}
}

@book{coleman2015,
	place={Cambridge},
	title={Introduction to Many-Body Physics},
	publisher={Cambridge University Press},
	author={Coleman, Piers},
	year={2015},
	url={https://doi.org/10.1017/CBO9781139020916}}

@article{bulla2008,
	title={Numerical renormalization group method for quantum impurity systems},
	author={Bulla, Ralf and Costi, Theo A and Pruschke, Thomas},
	journal={Rev. Mod. Phys.},
	volume={80},
	number={2},
	pages={395--450},
	year={2008},
	doi={10.1103/RevModPhys.80.395},
	url={https://doi.org/10.1103/RevModPhys.80.395},
	publisher={APS}
}

@article{clark2000andreevscattering,
  title = {Andreev scattering and the Kondo effect},
  author = {Clerk, Aashish A. and Ambegaokar, Vinay and Hershfield, Selman},
  journal = {Phys. Rev. B},
  volume = {61},
  issue = {5},
  pages = {3555--3562},
  numpages = {0},
  year = {2000},
  month = {Feb},
  publisher = {American Physical Society},
  doi = {10.1103/PhysRevB.61.3555},
  url = {https://link.aps.org/doi/10.1103/PhysRevB.61.3555}
}

@article{law2009majorana,
  title = {Majorana Fermion Induced Resonant Andreev Reflection},
  author = {Law, K. T. and Lee, Patrick A. and Ng, T. K.},
  journal = {Phys. Rev. Lett.},
  volume = {103},
  issue = {23},
  pages = {237001},
  numpages = {4},
  year = {2009},
  month = {Dec},
  publisher = {American Physical Society},
  doi = {10.1103/PhysRevLett.103.237001},
  url = {https://link.aps.org/doi/10.1103/PhysRevLett.103.237001}
}

@article{deacon2010kondo,
  title = {Kondo-enhanced Andreev transport in single self-assembled InAs quantum dots contacted with normal and superconducting leads},
  author = {Deacon, R. S. and Tanaka, Y. and Oiwa, A. and Sakano, R. and Yoshida, K. and Shibata, K. and Hirakawa, K. and Tarucha, S.},
  journal = {Phys. Rev. B},
  volume = {81},
  issue = {12},
  pages = {121308(R)},
  numpages = {4},
  year = {2010},
  month = {Mar},
  publisher = {American Physical Society},
  doi = {10.1103/PhysRevB.81.121308},
  url = {https://link.aps.org/doi/10.1103/PhysRevB.81.121308}
}

@article{pillet2013tunneling,
  title = {Tunneling spectroscopy of a single quantum dot coupled to a superconductor: From Kondo ridge to Andreev bound states},
  author = {Pillet, J.-D. and Joyez, P. and \ifmmode \check{Z}\else \v{Z}\fi{}itko, Rok and Goffman, M. F.},
  journal = {Phys. Rev. B},
  volume = {88},
  issue = {4},
  pages = {045101},
  numpages = {6},
  year = {2013},
  month = {Jul},
  publisher = {American Physical Society},
  doi = {10.1103/PhysRevB.88.045101},
  url = {https://link.aps.org/doi/10.1103/PhysRevB.88.045101}
}

@article{fidkowski2012universal,
  title = {Universal transport signatures of Majorana fermions in superconductor-Luttinger liquid junctions},
  author = {Fidkowski, Lukasz and Alicea, Jason and Lindner, Netanel H. and Lutchyn, Roman M. and Fisher, Matthew P. A.},
  journal = {Phys. Rev. B},
  volume = {85},
  issue = {24},
  pages = {245121},
  numpages = {22},
  year = {2012},
  month = {Jun},
  publisher = {American Physical Society},
  doi = {10.1103/PhysRevB.85.245121},
  url = {https://link.aps.org/doi/10.1103/PhysRevB.85.245121}
}

@article{affleck2013topological,
	doi = {10.1088/1742-5468/2013/06/P06011},
	url = {https://doi.org/10.1088/1742-5468/2013/06/P06011},
	year = {2013},
	month = {jun},
	publisher = {IOP Publishing and SISSA},
	volume = {2013},
	number = {06},
	pages = {P06011},
	author = {Affleck, Ian and Giuliano, Domenico},
	title = {Topological superconductor–Luttinger liquid junctions},
	journal = {Journal of Statistical Mechanics: Theory and Experiment}
}

@article{white1992density,
  title = {Density matrix formulation for quantum renormalization groups},
  author = {White, Steven R.},
  journal = {Phys. Rev. Lett.},
  volume = {69},
  issue = {19},
  pages = {2863--2866},
  numpages = {0},
  year = {1992},
  month = {Nov},
  publisher = {American Physical Society},
  doi = {10.1103/PhysRevLett.69.2863},
  url = {https://link.aps.org/doi/10.1103/PhysRevLett.69.2863}
}

@article{white1992real,
  title = {Real-space quantum renormalization groups},
  author = {White, S. R. and Noack, R. M.},
  journal = {Phys. Rev. Lett.},
  volume = {68},
  issue = {24},
  pages = {3487--3490},
  numpages = {0},
  year = {1992},
  month = {Jun},
  publisher = {American Physical Society},
  doi = {10.1103/PhysRevLett.68.3487},
  url = {https://link.aps.org/doi/10.1103/PhysRevLett.68.3487}
}

@article{schollwock2005density,
  title = {The density-matrix renormalization group},
  author = {Schollw\"ock, U.},
  journal = {Rev. Mod. Phys.},
  volume = {77},
  issue = {1},
  pages = {259--315},
  numpages = {0},
  year = {2005},
  month = {Apr},
  publisher = {American Physical Society},
  doi = {10.1103/RevModPhys.77.259},
  url = {https://link.aps.org/doi/10.1103/RevModPhys.77.259}
}

@article{schollwock2011density,
title = {The density-matrix renormalization group in the age of matrix product states},
journal = {Annals of Physics},
volume = {326},
number = {1},
pages = {96-192},
year = {2011},
note = {January 2011 Special Issue},
issn = {0003-4916},
doi = {https://doi.org/10.1016/j.aop.2010.09.012},
url = {https://www.sciencedirect.com/science/article/pii/S0003491610001752},
author = {Ulrich Schollw\"ock},
}

@article{golub2011kondo,
  title = {Kondo Correlations and Majorana Bound States in a Metal to Quantum-Dot to Topological-Superconductor Junction},
  author = {Golub, A. and Kuzmenko, I. and Avishai, Y.},
  journal = {Phys. Rev. Lett.},
  volume = {107},
  issue = {17},
  pages = {176802},
  numpages = {5},
  year = {2011},
  month = {Oct},
  publisher = {American Physical Society},
  doi = {10.1103/PhysRevLett.107.176802},
  url = {https://link.aps.org/doi/10.1103/PhysRevLett.107.176802}
}

@article{lee2013kondo,
	title={Kondo effect in a quantum dot side-coupled to a topological superconductor},
	author={Lee, Minchul and Lim, Jong Soo and L{\'o}pez, Rosa},
	journal={Phys. Rev. B},
	volume={87},
	number={24},
	pages={241402},
	year={2013},
	doi={10.1103/PhysRevB.87.241402},
	url={https://doi.org/10.1103/PhysRevB.87.241402},
	publisher={APS}
}

@article{cheng2014interplay,
	title={Interplay between Kondo and Majorana interactions in quantum dots},
	author={Cheng, Meng and Becker, Michael and Bauer, Bela and Lutchyn, Roman M},
	journal={Phys. Rev. X},
	volume={4},
	number={3},
	pages={031051},
	year={2014},
	doi={10.1103/PhysRevX.4.031051},
	url={https://doi.org/10.1103/PhysRevX.4.031051},
	publisher={APS}
}

@article{vernek2014subtle,
	title={Subtle leakage of a Majorana mode into a quantum dot},
	author={Vernek, E and Penteado, PH and Seridonio, AC and Egues, Jos{\'e} Carlos},
	journal={Phys. Rev. B},
	volume={89},
	number={16},
	pages={165314},
	year={2014},
	doi={10.1103/PhysRevB.89.165314},
	url={https://doi.org/10.1103/PhysRevB.89.165314},
	publisher={APS}
}

@article{silva2020robustness,
	title={Robustness of the Kondo effect in a quantum dot coupled to Majorana zero modes},
	author={Silva, Joelson F and Da Silva, Luis GGV Dias and Vernek, E},
	journal={Phys. Rev. B},
	volume={101},
	number={7},
	pages={075428},
	year={2020},
	doi={10.1103/PhysRevB.101.075428},
	url={https://doi.org/10.1103/PhysRevB.101.075428},
	publisher={APS}
}

@article{majek2022majorana,
	title={Majorana-Kondo competition in a cross-shaped double quantum dot-topological superconductor system},
	author={Majek, Piotr and Weymann, Ireneusz},
	journal={J. Magn. Magn. Mater.},
	volume={549},
	pages={168935},
	year={2022},
	doi={10.1016/j.jmmm.2021.168935},
	url={https://doi.org/10.1016/j.jmmm.2021.168935},
	publisher={Elsevier}
}

@article{shen2025,
	title={Spin-charge-entangled Kondo effect induced by a side-coupled Majorana zero mode},
	author={Shen, Haojie and Su, Wei and Chen, MN and Wang, Xiaoqun},
	journal={Phys. Rev. B},
	volume={112},
	number={8},
	pages={085123},
	year={2025},
	doi={10.1103/kmpz-3ytk},
	url={https://doi.org/10.1103/kmpz-3ytk},
	publisher={APS}
}

@article{tang2025twochannel,
  title = {Two-channel Kondo behavior in the quantum XX chain with a boundary defect},
  author = {Tang, Yicheng and Kattel, Pradip and Pixley, J. H. and Andrei, Natan},
  journal = {Phys. Rev. B},
  volume = {112},
  issue = {2},
  pages = {L020303},
  numpages = {6},
  year = {2025},
  month = {Jul},
  publisher = {American Physical Society},
  doi = {10.1103/t78z-pyfr},
  url = {https://link.aps.org/doi/10.1103/t78z-pyfr}
}

@article{weichselbaum2012tensor,
	title={Tensor networks and the numerical renormalization group},
	author={Weichselbaum, Andreas},
	journal={Phys. Rev. B},
	volume={86},
	number={24},
	pages={245124},
	year={2012},
	doi={10.1103/PhysRevB.86.245124},
	url={https://doi.org/10.1103/PhysRevB.86.245124},
	publisher={APS}
}

@article{merker2012fdm,
  title = {Full density-matrix numerical renormalization group calculation of impurity susceptibility and specific heat of the Anderson impurity model},
  author = {Merker, L. and Weichselbaum, A. and Costi, T. A.},
  journal = {Phys. Rev. B},
  volume = {86},
  issue = {7},
  pages = {075153},
  numpages = {8},
  year = {2012},
  month = {Aug},
  publisher = {American Physical Society},
  doi = {10.1103/PhysRevB.86.075153},
  url = {https://link.aps.org/doi/10.1103/PhysRevB.86.075153}
}

@article{costi2000kondo,
  title={Kondo effect in a magnetic field and the magnetoresistivity of Kondo alloys},
  author={Costi, TA},
  journal={Phys. Rev. Lett.},
  volume={85},
  number={7},
  pages={1504--1507},
  year={2000},
  doi={10.1103/PhysRevLett.85.1504},
  url={https://doi.org/10.1103/PhysRevLett.85.1504},
  publisher={APS}
}

@inbook{affleck2010kondo,
	author = {Ian Affleck},
	title = {The {Kondo} Screening Cloud: What It Is and How to Observe It},
	booktitle = {Perspectives of Mesoscopic Physics},
	pages = {1-44},
	doi = {10.1142/9789814299442_0001},
	URL = {https://www.worldscientific.com/doi/abs/10.1142/9789814299442_0001},
}

@article{park2013directly,
  title={How to directly measure a Kondo cloud’s length},
  author={Park, Jinhong and Lee, S-SB and Oreg, Yuval and Sim, H-S},
  journal={Phys. Rev. Lett.},
  volume={110},
  number={24},
  pages={246603},
  year={2013},
  doi={10.1103/PhysRevLett.110.246603},
  url={https://doi.org/10.1103/PhysRevLett.110.246603},
  publisher={APS}
}

@article{ivan2020observation,
  title={Observation of the Kondo screening cloud},
  author={Borzenets, Ivan V. and Shim, Jeongmin and Chen, Jason C. H. and Ludwig, Arne and Wieck, Andreas D. and Tarucha, Seigo and Sim, H.-S. and Yamamoto, Michihisa},
  journal={Nature},
  volume={579},
  number={7798},
  pages={210--213},
  year={2020},
  doi={10.1038/s41586-020-2058-6},
  url={https://doi.org/10.1038/s41586-020-2058-6},
  publisher={Nature Publishing Group UK London}
}

@article{bollmann2024topological,
  title = {Topological Kondo effect with spinful Majorana fermions},
  author = {Bollmann, Steffen and V\"ayrynen, Jukka I. and K\"onig, Elio J.},
  journal = {Phys. Rev. B},
  volume = {110},
  issue = {3},
  pages = {035136},
  numpages = {23},
  year = {2024},
  month = {Jul},
  publisher = {American Physical Society},
  doi = {10.1103/PhysRevB.110.035136},
  url = {https://link.aps.org/doi/10.1103/PhysRevB.110.035136}
}

@article{nishimoto2004density,
doi = {10.1088/0953-8984/16/4/010},
url = {https://doi.org/10.1088/0953-8984/16/4/010},
year = {2004},
month = {jan},
publisher = {},
volume = {16},
number = {4},
pages = {613},
author = {S Nishimoto and E Jeckelmann},
title = {Density-matrix renormalization group approach to quantum impurity problems},
journal = {Journal of Physics: Condensed Matter},
}

@article{hand2006spinrelations,
  title = {Spin Correlations and Finite-Size Effects in the One-Dimensional Kondo Box},
  author = {Hand, Thomas and Kroha, Johann and Monien, Hartmut},
  journal = {Phys. Rev. Lett.},
  volume = {97},
  issue = {13},
  pages = {136604},
  numpages = {4},
  year = {2006},
  month = {Sep},
  publisher = {American Physical Society},
  doi = {10.1103/PhysRevLett.97.136604},
  url = {https://link.aps.org/doi/10.1103/PhysRevLett.97.136604}
}

@article{holzner2009kondo,
  title = {Kondo screening cloud in the single-impurity Anderson model: A density matrix renormalization group study},
  author = {Holzner, Andreas and McCulloch, Ian P. and Schollw\"ock, Ulrich and von Delft, Jan and Heidrich-Meisner, Fabian},
  journal = {Phys. Rev. B},
  volume = {80},
  issue = {20},
  pages = {205114},
  numpages = {8},
  year = {2009},
  month = {Nov},
  publisher = {American Physical Society},
  doi = {10.1103/PhysRevB.80.205114},
  url = {https://link.aps.org/doi/10.1103/PhysRevB.80.205114}
}

@article{kattel2026thermodynamics,
  title = {Thermodynamics in a split Hilbert space: Quantum impurity at the edge of a one-dimensional superconductor},
  author = {Kattel, Pradip and Zhakenov, Abay and Andrei, Natan},
  journal = {Phys. Rev. B},
  volume = {113},
  issue = {19},
  pages = {195155},
  numpages = {21},
  year = {2026},
  month = {May},
  publisher = {American Physical Society},
  doi = {10.1103/qw2d-k8wv},
  url = {https://link.aps.org/doi/10.1103/qw2d-k8wv}
}

@article{luh1965,
	title={Bound state in superconductors with paramagnetic impurities},
	author={Luh, Y.},
	journal={Acta Phys. Sin.},
	volume={21},
	pages={75},
	year={1965},
	doi={10.7498/APS.21.75},
	url={https://doi.org/10.7498/APS.21.75},
	publisher={Chinese Physical Society}
}

@article{shiba1968classical,
	title={Classical spins in superconductors},
	author={Shiba, Hiroyuki},
	journal={Prog. Theor. Phys.},
	volume={40},
	number={3},
	pages={435--451},
	year={1968},
	doi={10.1143/PTP.40.435},
	url={https://doi.org/10.1143/PTP.40.435},
	publisher={Oxford University Press}
}

@article{rusinov1969superconductivity,
	title={Superconductivity near a paramagnetic impurity},
	author={Rusinov, A. I.},
	journal={Sov. J. Exp. Theor. Phys. Lett.},
	volume={9},
	pages={85--87},
	year={1969}
}

@article{pasnoori2020kondo,
	title={Kondo impurity at the edge of a superconducting wire},
	author={Pasnoori, Parameshwar R. and Rylands, Colin and Andrei, Natan},
	journal={Phys. Rev. Research},
	volume={2},
	number={1},
	pages={013006},
	year={2020},
	doi={10.1103/PhysRevResearch.2.013006},
	url={https://doi.org/10.1103/PhysRevResearch.2.013006},
	publisher={APS}
}

@article{pasnoori2022rise,
	title={Rise and fall of Yu-Shiba-Rusinov bound states in charge-conserving $s$-wave one-dimensional superconductors},
	author={Pasnoori, Parameshwar R. and Andrei, Natan and Rylands, Colin and Azaria, Patrick},
	journal={Phys. Rev. B},
	volume={105},
	number={17},
	pages={174517},
	year={2022},
	doi={10.1103/PhysRevB.105.174517},
	url={https://doi.org/10.1103/PhysRevB.105.174517},
	publisher={APS}
}

@article{moca2025fractionalized,
	title={Spectral properties of fractionalized Shiba states},
	author={Moca, Catalin Pascu and Hajdu, Csanad and Dora, Balazs and Zarand, Gergely},
	journal={Phys. Rev. Lett.},
	volume={135},
	pages={126502},
	year={2025},
	doi={10.1103/g4sp-t82d},
	url={https://doi.org/10.1103/g4sp-t82d},
	publisher={APS}
}

@article{moca2021kondo,
	title={Kondo cloud in a superconductor},
	author={Moca, Catalin Pascu and Weymann, Ireneusz and Werner, Miklos Antal and Zarand, Gergely},
	journal={Phys. Rev. Lett.},
	volume={127},
	number={18},
	pages={186804},
	year={2021},
	doi={10.1103/PhysRevLett.127.186804},
	url={https://doi.org/10.1103/PhysRevLett.127.186804},
	publisher={APS}
}

@article{huang2025largeN,
	title={A large-$N$ approach to magnetic impurities in superconductors},
	author={Huang, Chen-How and Lobos, Alejandro M. and Cazalilla, Miguel A.},
	journal={SciPost Phys.},
	volume={18},
	number={3},
	pages={087},
	year={2025},
	doi={10.21468/SciPostPhys.18.3.087},
	url={https://doi.org/10.21468/SciPostPhys.18.3.087},
	publisher={SciPost Foundation}
}

@misc{appendix,
	title={See appendices for more details.},
}

@article{prada2012transport,
	title={Transport spectroscopy of NS nanowire junctions with Majorana fermions},
	author={Prada, Elsa and San-Jose, Pablo and Aguado, Ramon},
	journal={Phys. Rev. B},
	volume={86},
	pages={180503},
	year={2012},
	doi={10.1103/PhysRevB.86.180503},
	url={https://doi.org/10.1103/PhysRevB.86.180503},
	publisher={APS}
}

@article{dassarma2012splitting,
	title={Splitting of the zero-bias conductance peak as smoking gun evidence for the existence of the Majorana mode in a superconductor-semiconductor nanowire},
	author={Das Sarma, Sankar and Sau, Jay D. and Stanescu, Tudor D.},
	journal={Phys. Rev. B},
	volume={86},
	pages={220506},
	year={2012},
	doi={10.1103/PhysRevB.86.220506},
	url={https://doi.org/10.1103/PhysRevB.86.220506},
	publisher={APS}
}

@article{jones1988twoimpurity,
	title={Low-temperature properties of the two-impurity Kondo Hamiltonian},
	author={Jones, B. A. and Varma, C. M. and Wilkins, J. W.},
	journal={Phys. Rev. Lett.},
	volume={61},
	pages={125--128},
	year={1988},
	doi={10.1103/PhysRevLett.61.125},
	url={https://doi.org/10.1103/PhysRevLett.61.125},
	publisher={APS}
}

@article{craig2004tunable,
	title={Tunable Nonlocal Spin Control in a Coupled-Quantum Dot System},
	author={Craig, N. J. and Taylor, J. M. and Lester, E. A. and Marcus, C. M. and Hanson, M. P. and Gossard, A. C.},
	journal={Science},
	volume={304},
	pages={565--567},
	year={2004},
	doi={10.1126/science.1095452},
	url={https://doi.org/10.1126/science.1095452},
	publisher={American Association for the Advancement of Science}
}

@article{zitko2006enhanced,
	title={Enhanced conductance through side-coupled double quantum dots},
	author={{\v Z}itko, Rok and Bon{\v c}a, Janez},
	journal={Phys. Rev. B},
	volume={73},
	pages={035332},
	year={2006},
	doi={10.1103/PhysRevB.73.035332},
	url={https://doi.org/10.1103/PhysRevB.73.035332},
	publisher={APS}
}

@article{deng2018nonlocality,
	title={Nonlocality of Majorana modes in hybrid nanowires},
	author={Deng, M.-T. and Vaitiek{\.e}nas, S. and Prada, E. and San-Jose, P. and Nyg{\aa}rd, J. and Krogstrup, P. and Aguado, R. and Marcus, C. M.},
	journal={Phys. Rev. B},
	volume={98},
	pages={085125},
	year={2018},
	doi={10.1103/PhysRevB.98.085125},
	url={https://doi.org/10.1103/PhysRevB.98.085125},
	publisher={APS}
}

@article{bork2011twoimpurity,
	title={A tunable two-impurity Kondo system in an atomic point contact},
	author={Bork, Jakob and Zhang, Yong-Hui and Diekh{\"o}ner, Lars and Borda, L{\'a}szl{\'o} and Simon, Pascal and Kroha, Johann and Wahl, Peter and Kern, Klaus},
	journal={Nat. Phys.},
	volume={7},
	number={11},
	pages={901--906},
	year={2011},
	doi={10.1038/nphys2076},
	url={https://doi.org/10.1038/nphys2076},
	publisher={Nature Publishing Group}
}

@article{sasaki2006nonlocal,
	title={Nonlocal control of the Kondo effect in a double quantum dot--quantum wire coupled system},
	author={Sasaki, S. and Kang, S. and Kitagawa, K. and Yamaguchi, M. and Miyashita, S. and Maruyama, T. and Tamura, H. and Akazaki, T. and Hirayama, Y. and Takayanagi, H.},
	journal={Phys. Rev. B},
	volume={73},
	pages={161303},
	year={2006},
	doi={10.1103/PhysRevB.73.161303},
	url={https://doi.org/10.1103/PhysRevB.73.161303},
	publisher={APS}
}

@article{deng2016coupledqd,
	title={Majorana bound state in a coupled quantum-dot hybrid-nanowire system},
	author={Deng, M. T. and Vaitiek{\.e}nas, S. and Hansen, E. B. and Danon, J. and Leijnse, M. and Flensberg, K. and Nyg{\aa}rd, J. and Krogstrup, P. and Marcus, C. M.},
	journal={Science},
	volume={354},
	number={6319},
	pages={1557--1562},
	year={2016},
	doi={10.1126/science.aaf3961},
	url={https://doi.org/10.1126/science.aaf3961},
	publisher={American Association for the Advancement of Science}
}

@article{emery1992mapping,
	title={Mapping of the two-channel Kondo problem to a resonant-level model},
	author={Emery, V. J. and Kivelson, Steven},
	journal={Phys. Rev. B},
	volume={46},
	number={17},
	pages={10812--10817},
	year={1992},
	doi={10.1103/PhysRevB.46.10812},
	url={https://doi.org/10.1103/PhysRevB.46.10812},
	publisher={APS}
}

@article{lopes2020anyons,
	title={Anyons in multichannel Kondo systems},
	author={Lopes, Pedro L. S. and Affleck, Ian and Sela, Eran},
	journal={Phys. Rev. B},
	volume={101},
	number={8},
	pages={085141},
	year={2020},
	doi={10.1103/PhysRevB.101.085141},
	url={https://doi.org/10.1103/PhysRevB.101.085141},
	publisher={APS}
}

@article{kattel2024overscreened,
	title={Overscreened spin-1/2 Kondo impurity and Shiba state at the edge of a one-dimensional spin-1 superconducting wire},
	author={Kattel, Pradip and Zhakenov, Abay and Andrei, Natan},
	journal={arXiv preprint arXiv:2412.01924},
	year={2024},
	eprint={2412.01924},
	archivePrefix={arXiv},
	primaryClass={cond-mat.str-el},
	doi={10.48550/arXiv.2412.01924},
	url={https://arxiv.org/abs/2412.01924}
}

@article{affleck1995conformal,
	title={Conformal-field-theory approach to the two-impurity Kondo problem: Comparison with numerical renormalization-group results},
	author={Affleck, Ian and Ludwig, Andreas W. W. and Jones, Barbara A.},
	journal={Phys. Rev. B},
	volume={52},
	number={13},
	pages={9528--9546},
	year={1995},
	doi={10.1103/PhysRevB.52.9528},
	url={https://doi.org/10.1103/PhysRevB.52.9528},
	publisher={APS}
}

	\newpage
	\appendix
	\begin{widetext}
	
	\section{Schrieffer-Wolff Transformation}
	\subsection{Symmetric case}
	In this section we perform the Schrieffer-Wolff (SW) transformation of the MZM-QD subsystem for the \(\mathbb{Z}_2\times {\rm SU}_{\bf L}(2)\) symmetric case. By doing this, the starting point of this work, i.e., Eq. (\ref{eq:heff1}-~\ref{eq:heff2}), is derived from an Anderson-type impurity Hamiltonian
	\begin{equation}
		H = \sum_{\sigma} \epsilon_d d_{\sigma}^\dagger d_{\sigma} + U d_{\uparrow}^\dagger d_{\uparrow} d_{\downarrow}^\dagger d_{\downarrow} + \sum_{{\bf k}\sigma}V_{\bf k} (d_{\sigma}^\dagger c_{{\bf k}\sigma} + h.c.) + \sum_{\bf k} \epsilon_{\bf k} c_{{\bf k}\sigma}^\dagger c_{{\bf k}\sigma} + \frac{\lambda}{2} \gamma_{-1}(d_{\uparrow} - d_{\uparrow}^\dagger) , \label{eq:a1:anderson}
	\end{equation}
	For the moment, the effect of the MZM \(\gamma_{-2}\) is dropped and the local MZM-QD Hamiltonian reads
	\begin{equation}
		H_{\gamma_{-1}d} = \frac{\lambda}{2} \gamma_{-1}(d_{\uparrow} - d_{\uparrow}^{\dagger}) + \sum_{\sigma} \epsilon_{d} n_{d\sigma} + U d_{\uparrow}^\dagger d_{\uparrow} d_{\downarrow}^\dagger d_{\downarrow}
	\end{equation}
	The spectrum of the subsystem is
	\begin{equation}
		\begin{array}{lll}
			E^{\pm}_{(0,+)} = \frac{1}{2}(\epsilon_d \pm \Delta_E), & \quad \ket{0,+}^{\pm} = (\epsilon_d \mp \Delta_E) \ket{0,0} &+\ \lambda\ket{1,\uparrow}, \\
			E^{\pm}_{(0,-)} = \frac{1}{2}(\epsilon_d^+ \pm \tilde{\Delta}_E), & \quad \ket{0,-}^{\pm} = (\epsilon_d^- \mp \tilde{\Delta}_E) \ket{0,\uparrow\downarrow} &-\ \lambda \ket{1,\downarrow}, \\
			E^{\pm}_{(1,+)} = \frac{1}{2}(\epsilon_d \pm \Delta_E), & \quad \ket{1,+}^{\pm} = (\epsilon_d \pm \Delta_E) \ket{1,0} &+\ \lambda \ket{0,\uparrow}, \\
			E^{\pm}_{(1,-)} = \frac{1}{2}(\epsilon_d^+ \pm \tilde{\Delta}_E), & \quad \ket{1,-}^{\pm} = (\epsilon_d^- \pm \tilde{\Delta}_E) \ket{1,\uparrow\downarrow} &-\ \lambda \ket{0,\downarrow},
		\end{array}
	\end{equation}
	where \(\Delta_E = \sqrt{\epsilon_d^2+\lambda^2}\), \(\tilde{\Delta}_E = \sqrt{(\epsilon_d-\delta U)^2+\lambda^2}\), \(\epsilon_d^{\pm}=\epsilon_d \pm \delta U\), and \(\delta U=U+2\epsilon_d\) evaluates the deviation from particle-hole symmetry. The eigenstates are labeled by the \(\mathbbm{Z}_2 \times {\rm SU}_{\bf L}(2)\) symmetry \((P,L^z)\), and so are the corresponding eigenenergies \(E_{(P_f,L^z)}\). The normalization factors \(\mathcal{N}\) are omitted for brevity. 
	
	We see the MZM-QD subsystem has an obvious parity degeneracy of two since \(E_{(P_f,L^z)}=E_{(1-P_f,L^z)}\). Generally the eigenenergies of different \({\rm SU}_{\bf L}(2)\) indices are different, because the pseudospin \({\bf L}\) consists of charge degrees of freedom. 
	At the charge-symmetric point, \(\epsilon_d^{\pm}=\epsilon_{d}\), \(\tilde{\Delta}_E=\Delta_E\), and the spectrum reduces to a fourfold degeneracy. There are only two different levels, \(E^\pm=\frac{1}{2}(\epsilon_d \pm \Delta_E)\), so that the Hilbert space of the subsystem can be divided into a high- and a low-energy subspace, denoted by \(\mathcal{H}^+\) and \(\mathcal{H}^-\). Assuming the hybridization between the QD and the NL bath, \(V_{\bf k}\), to be much weaker than the energy gap \(\Delta_E\) between the two subspaces, the SW transformation can be applied to project out the high-energy states and yield a low-energy effective Hamiltonian
	\begin{equation}
		H_{\mathrm{eff}}=H^{--} + H^{-+} \frac{1}{E-H^{++}} H^{+-}, \label{eq:a1:sw}
	\end{equation}
	where \(H^{\alpha\beta} = \sum_{m\in\alpha,n\in\beta} P_{m} H P_{n}\), and \(P_{n} = \ket{n}\bra{n}\) is the projection operator onto eigenstates of the high- or low-energy subspaces, with \(\alpha,\beta=\pm\). 
	The first term \(H^{--}\) contains the real processes, namely the transitions within the low-energy states,
	\begin{equation}
		H^{--}=A\left(\ket{1,+}^-,\ket{0,+}^-,\ket{1,-}^-,\ket{0,-}^-\right)
		\begin{pmatrix} 
			0 & c_{0\uparrow}-c_{0\uparrow}^{\dagger} & 0 & -2 c_{0\downarrow}^{\dagger}\\
			c_{0\uparrow}^{\dagger}-c_{0\uparrow} & 0 & 2 c_{0\downarrow}^{\dagger} & 0\\
			0 & 2 c_{0\downarrow} & 0 & c_{0\uparrow}^{\dagger}-c_{0\uparrow}\\
			-2 c_{0\downarrow} & 0 & c_{0\uparrow}-c_{0\uparrow}^{\dagger} & 0
		\end{pmatrix}
		\left(\begin{array}{c}
			^-\bra{1,+}\\
			^-\bra{0,+}\\
			^-\bra{1,-}\\
			^-\bra{0,-}
		\end{array}\right),
	\end{equation} 
	where \(c_{0\sigma}^{(\dagger)}=\sum_{\bf k}c_{{\bf k}\sigma}^{(\dagger)}\) is the electron annihilation (creation) operator at the QD point, and \(A=4 V \lambda (\epsilon_d-\Delta_E) / \mathcal{N}\). Using the \({\rm SU}_{\bf L}(2)\) operators \(L_+\), \(L_-\), and \(L^z\), the Hamiltonian can be written in a compact form
	\begin{equation}
		\begin{aligned}
			H^{--} &= J_M \gamma_{-1}\left[ L^z (c_{0\uparrow}-c_{0\uparrow}^{\dagger}) + L_+ (-c_{0\downarrow}^{\dagger}) + L_- c_{0\downarrow} \right]\\
			&= J_M i\gamma_{-1} {\bf L} \cdot {\bm \xi}_{0},
		\end{aligned}
	\end{equation}
	where \({\bf L}={\bf S}+{\bm \eta}\) is the SU(2) pseudospin of the QD containing both spin and charge degrees of freedom, with
	\begin{gather}
		S^z = \frac{1}{2}(d_\uparrow^\dagger d_\uparrow-d_\downarrow^\dagger d_\downarrow),\ S_+ = d_\uparrow^\dagger d_\downarrow,\ S_- = d_\downarrow^\dagger d_\uparrow,\\
		\eta^z = \frac{1}{2}(1-\sum_{\sigma} d_\sigma^\dagger d_\sigma),\ \eta_+ = -d_\downarrow d_\uparrow,\ \eta_- = -d_\uparrow^\dagger d_\downarrow^\dagger,
	\end{gather}
	and 
	\begin{equation}
		\xi^z_0 = i(c_{0\uparrow}^\dagger-c_{0\uparrow}),\ \xi^+_0 = c_{0\downarrow},\ \xi^-_0 = -c_{0\uparrow}^\dagger
	\end{equation}
	which satisfy the Clifford algebra \(\{\xi_{i\mu},\xi_{j\nu}\}=2\delta_{ij}\delta_{\mu\nu}\), and \(c_{0\sigma}^{(\dagger)}\) is the annihilation (creation) operator of the electron at the impurity location. We have employed a constant hybridization factor \(V_{\bf k}\equiv V\), and the effective interaction is to first order in \(V\)
	\begin{equation}
		J_M = \frac{\lambda}{\lambda - \Delta_E}V \xrightarrow{\lambda\to0} \frac{\lambda}{|\epsilon_d|}V.
	\end{equation}
	
	The second term in Eq. (\ref{eq:a1:sw}) contains the virtual processes mediated by instantaneous transitions to higher energy states. Up to the second order of \(V\), we have
	\begin{equation}
		H^{-+} \frac{1}{E-H^{++}} H^{+-} = J_K {\bf L} \cdot {\bf s}_{0}+J_Q {\bf L} \cdot {\bm \eta}_{0} + H_{\rm bath},
	\end{equation}
	where
	\begin{align}
		J_K &= 2V^{2}\frac{\epsilon_{d}(\epsilon_{d}-\Delta_E)}{\left(\epsilon_s^2+\lambda^2\right)^{3/2}} \xrightarrow{\lambda\to0} \frac{V^2}{-\epsilon_d}, \\
		J_Q &= 2V^{2}\frac{\epsilon_d(\epsilon_d+\Delta_E)}{\left(\epsilon_d^2+\lambda^2\right)^{3/2}} \xrightarrow{\lambda\to0} 2V^2 \left(\frac{1}{\epsilon_d}+\frac{1}{|\epsilon_{d}|}\right),
	\end{align}
	and \({\bf s}_0\) and \({\bm \eta}_0\) are spin and charge SU(2) operators of the electron \(c_0\), respectively. 
	
	After projecting out the high-energy Hilbert space \(\mathcal{H}^+\), the charge fluctuation on the QD is frozen, and the final effective Hamiltonian is
	\begin{equation}
		H_{\rm eff}=J_M i \gamma_{-1} {\bf L} \cdot {\bm \xi}_{0} + J_K {\bf L} \cdot {\bf s}_{0} + J_Q {\bf L} \cdot {\bm \eta}_{0}. 
	\end{equation}
	In the large positive \(U\) limit, the charge DoFs are frozen on the QD, and the \(J_Q\) term vanishes. 
	
	\subsection{\texorpdfstring{$\delta U$: ${\rm SU}_{\bf L}(2)$ asymmetry}{delta U: SU L(2) asymmetry}}
	For \(U\neq2\epsilon_d\), the violation of the particle-hole symmetry can be evaluated by \(H_U=\delta U d_{\uparrow}^\dagger d_{\uparrow} d_{\downarrow}^\dagger d_{\downarrow}\) with \(\delta U=U+2\epsilon_d\). Inside the low-energy Hilbert subspace \(\mathcal{H}^-\), we have
	\begin{equation}
		\begin{array}{l}
			H_U \ket{0,+}^- = 0, \\
			H_U \ket{0,-}^- = \lambda \delta U \ket{0,\uparrow\downarrow}, \\
			H_U \ket{1,+}^- = 0, \\
			H_U \ket{1,-}^- = -\lambda \delta U \ket{1,\uparrow\downarrow}.
		\end{array}
	\end{equation}
	After performing the inner product for each real process, we get the resulting Hamiltonian formulated as 
	\begin{equation}
		H_U = \frac{\lambda^2}{\Delta_E^2-\epsilon_d\Delta_E} \delta U (L^z-\frac{1}{2}),
	\end{equation}
	which represents an effective Zeeman field \(h=\frac{\lambda^2}{\Delta_E^2-\epsilon_d\Delta_E} \delta U\) on the local pseudospin \({\bf L}_{\rm QD}\). In the effective Hamiltonian Eq. (\ref{eq:heff1}-~\ref{eq:heff2}), this corresponds to a \({\rm SU}_{\bf L}(2)\) symmetry breaking.

	\subsection{\texorpdfstring{$\gamma_{-2}$: $\mathbb{Z}_2$ asymmetry}{gamma(-2): Z2 asymmetry}}	
	When the overlap of MZMs is considered, the \(\mathbbm{Z}_2\) parity symmetry is broken and the subsystem Hamiltonian reads
	\begin{equation}
		H_{\gamma_{-2}\gamma_{-1}d} = \epsilon_m i \gamma_{-1}\gamma_{-2} + \frac{\lambda}{2} \gamma_{-1}(d_{\uparrow} - d_{\uparrow}^{\dagger}) + \frac{\mu}{2} i \gamma_{-2}(d_{\uparrow} + d_{\uparrow}^\dagger) + \sum_{\sigma} \epsilon_{d} n_{d\sigma} + U d_{\uparrow}^\dagger d_{\uparrow} d_{\downarrow}^\dagger d_{\downarrow}
	\end{equation}
	The spectrum of the subsystem with the third term ignored is
	\begin{equation}
		\begin{array}{lll}
			E^{\pm}_{0,+} = \frac{1}{2}(\epsilon_d + 2\epsilon_m \pm \Delta_E^+), & \quad \ket{0,+}^{\pm} = (\epsilon_d + 2\epsilon_m \mp \Delta_E^+) \ket{0,0} &+\ \lambda\ket{1,\uparrow}, \\
			E^{\pm}_{0,-} = \frac{1}{2}(\epsilon_d^+ + 2\epsilon_m \pm \tilde{\Delta}_E^+), & \quad \ket{0,-}^{\pm} = (\epsilon_d^- + 2\epsilon_m \mp \tilde{\Delta}_E^+) \ket{0,\uparrow\downarrow} &-\ \lambda \ket{1,\downarrow}, \\
			E^{\pm}_{1,+} = \frac{1}{2}(\epsilon_d + 2\epsilon_m \pm \Delta_E^-), & \quad \ket{1,+}^{\pm} = (\epsilon_d - 2\epsilon_m \pm \Delta_E^-) \ket{1,0} &+\ \lambda \ket{0,\uparrow}, \\
			E^{\pm}_{1,-} = \frac{1}{2}(\epsilon_d^+ + 2\epsilon_m \pm \tilde{\Delta}_E^-), & \quad \ket{1,-}^{\pm} = (\epsilon_d^- - 2\epsilon_m \pm \tilde{\Delta}_E^-) \ket{1,\uparrow\downarrow} &-\ \lambda \ket{0,\downarrow},
		\end{array}
	\end{equation}
	where \(\Delta_E^{\pm} = \sqrt{(\epsilon_d \pm 2\epsilon_m)^2+\lambda^2}\), \(\tilde{\Delta}_E^{\pm} = \sqrt{(\epsilon_d \pm 2\epsilon_m - \delta U)^2+\lambda^2}\), \(\epsilon_d^{\pm}=\epsilon_d \pm \delta U\). The eigenstates are still labeled by \((P_f,L^z)\). If we consider \(\delta U=0\), the local \({\rm SU}_{\bf L}(2)\) symmetry will be preserved. We see the topological degeneracy broken when \(\epsilon_m\neq 0\) because \(E_{0,L^z}\neq E_{1,L^z}\).
	
	This asymmetry arises from the Majorana overlap term, which is retained in the Hamiltonian and is not modified by the SW transformation. The resulting Majorana-Kondo Hamiltonian is Eq.~(\ref{eq:heff1}-~\ref{eq:heff2}) in the main text
	\begin{equation}
		H_{\rm eff} = J_K \sum_{\bf kk'} {\bf L}\cdot{\bf S}_{\bf kk'} + J_Q \sum_{\bf kk'} {\bf L}\cdot{\bm \eta}_{\bf kk'} + J_M \gamma_{-1} \sum_{\bf k}{\bf L}\cdot{\bm \xi}_{\bf k} + \sum_{\bf k} \epsilon_{\bf k} c_{{\bf k}\sigma}^\dagger c_{{\bf k}\sigma} + \epsilon_m i \gamma_{-1}\gamma_{-2} + h L^z.
		\label{eq:a1:heff}
	\end{equation}
	
	Finally, we consider the hybridization term \(H_{\gamma_{-2}d}= i \gamma_{-2}(d_{\uparrow} + d_{\uparrow}^\dagger)\). We have
	\begin{equation}
		\begin{array}{ll}
			H_{\gamma_{-2}d} \ket{0,+}^- = -\ (\epsilon_d + 2\epsilon_m + \Delta_E^+) \ket{1,\uparrow} &+\ \lambda \ket{0,0}, \\
			H_{\gamma_{-2}d} \ket{0,-}^- = -\ (\epsilon_d^- + 2\epsilon_m + \tilde{\Delta}_E^+) \ket{1,\downarrow} &-\ \lambda \ket{0,\uparrow\downarrow}, \\
			H_{\gamma_{-2}d} \ket{1,+}^- = +\ (\epsilon_d - 2\epsilon_m - \Delta_E^-) \ket{0,\uparrow} &+\ \lambda \ket{1,0}, \\
			H_{\gamma_{-2}d} \ket{1,-}^- = +\ (\epsilon_d^- - 2\epsilon_m - \tilde{\Delta}_E^-) \ket{0,\downarrow} &-\ \lambda \ket{1,\uparrow\downarrow}.
		\end{array}
	\end{equation}
	It is easy to see that every matrix element of \(H_{\gamma_{-2}d}\) within the low-energy Hilbert subspace \(\mathcal{H}^-\) vanishes, since the two terms in each inner product cancel each other. As a result, there is no contribution from the real processes, and the effect caused by this hybridization is only up to second order in \(\mu\). Additionally, we should have \(\mu\ll \lambda\) unless the MZMs \(\gamma_{-1}\) and \(\gamma_{-2}\) have a very large wavefunction overlap, in which case the Majorana properties can hardly be maintained. In this way, we are free to neglect the perturbation from this term. 
	
	\section{Strong-Coupling Limit}
	In this section, we derive the effective Hamiltonian at the lowest-energy strong coupling fixed point from Eq. (\ref{eq:a1:heff}).
	For the symmetric configuration, the ground states are a parity doublet
	\begin{align}
		\ket{0} &= (J_K+\sqrt{J_K^2+J_M^2})(\ket{1,\uparrow,\downarrow} - \ket{1,\downarrow,\uparrow}) +  J_M(\ket{0,\downarrow,0}-\ket{0,\uparrow,\uparrow\downarrow}),\label{eq:a2:gs1} \\
		\ket{1} &= (J_K+\sqrt{J_K^2+J_M^2})(\ket{0,\downarrow,\uparrow} - \ket{0,\uparrow,\downarrow}) +  J_M(\ket{1,\downarrow,0} - \ket{1,\uparrow,\uparrow\downarrow}), \label{eq:a2:gs2}
	\end{align}
	up to a normalization factor. They are both \({\rm SU}_{\bf L}(2)\) singlets with eigenvalue \(-\frac{3}{4}(J_K+\sqrt{J_K^2+J_M^2})\), and we have
	\begin{equation}
	P_f \ket{0} = \ket{0}, \quad P_f \ket{1} = -\ket{1}.
	\end{equation}
	According to Ref.~\cite{shen2025}, the strong-coupling Hamiltonian reads
	\begin{equation}
		H_{\rm sc} = \frac{J_M}{2\sqrt{J_K^2+J_M^2}}i\tilde{\gamma}_0\zeta_1 + H_{\rm bath}
	\end{equation}
	where \(\tilde{\gamma}_0=J_M\zeta_0-J_K\gamma_{-1}\), and \(\zeta_0=c_{0\uparrow}+c_{0\uparrow}^\dagger\), \(\zeta_1=i(c_{1\uparrow}-c_{1\uparrow}^\dagger)\) are Majorana operators.
	
	We first consider a small \(h L^z\) term, which commutes with the parity operator, \([P_f, L^z] = 0\). The first order perturbation turns out
	\begin{equation}
		\braket{0|L^z|1} = \braket{0| P_f^{\dagger} P_f L^z P_f^{\dagger} P_f |1} = \braket{0| P_f^{\dagger} L^z P_f |1} = - \braket{0|L^z|1},
	\end{equation}
	thus we have \(\braket{0|L^z|1} = \braket{1|L^z|0} = 0\), which means a small Zeeman field will not affect the non-Fermi liquid strong-coupling fixed point at low temperatures. When \(h\) is large and cannot be handled as a perturbation, we instead take the magnetic scattering terms as perturbations, since the impurity is highly polarized in this case. We start from the Hamiltonian
	\begin{equation}
		H_{\rm pol} = h L^z + J_M i\gamma_{-1}L^z \xi_{0z} + J_K L^z S_{0z},
	\end{equation}
	whose ground states are also a parity doublet
	\begin{align}
		\ket{0}_{\rm pol} &= (J_K+\sqrt{J_K^2+J_M^2})\ket{0,\downarrow,0} - J_M \ket{1,\downarrow,\uparrow},\label{eq:a2:gs3} \\
		\ket{1}_{\rm pol} &= (J_K+\sqrt{J_K^2+J_M^2})\ket{0,\downarrow,\uparrow} - J_M\ket{1,\downarrow,0}, \label{eq:a2:gs4}
	\end{align}
	but not \({\rm SU}_{\bf L}(2)\) singlets any more. The ground state energy is \(-\frac{1}{2}(h+J_K+\sqrt{J_K^2+J_M^2})\). Consider the first order perturbation of \(H_{\rm mag}=\frac{1}{2}J_M i\gamma_{-1}(L_+\xi_{0-} + L_-\xi_{0+}) + \frac{1}{2}J_K(L_+ S_{0-} + L_- S_{0+})\) 
	\begin{equation}
		\begin{aligned}
			H_{\rm mag}\ket{0_{\rm pol}} &= J_M(2J_K+\sqrt{J_K^2+J_M^2})\ket{1,\uparrow,\downarrow} - J_M^2\ket{0,\uparrow,\uparrow\downarrow}, \\
			H_{\rm mag}\ket{1_{\rm pol}} &= (J_K+J_M)(J_K+\sqrt{J_K^2+J_M^2})\ket{1,\uparrow,\uparrow\downarrow} - J_M^2\ket{0,\uparrow,\downarrow},
		\end{aligned}
	\end{equation}
	and we have \(\braket{1_{\rm pol}|H_{\rm mag}|0_{\rm pol}}=\braket{0_{\rm pol}|H_{\rm mag}|1_{\rm pol}}=0\).	
	In this way, we prove that the twofold ground state degeneracy is unaffected by the \({\rm SU}_{\bf L}(2)\) symmetry breaking in both the small- and large-\(h\) limits. We note that the properties of the ground states in the two cases are completely different. For a small Zeeman field perturbation, the ground states (\ref{eq:a2:gs1},~\ref{eq:a2:gs2}) are still \({\rm SU}_{\bf L}(2)\) singlets. For large \(h\), they cross over to the \({\bf L}\)-polarized states (\ref{eq:a2:gs3},~\ref{eq:a2:gs4}). The parity symmetry is conserved throughout this crossover, and the lowest-energy fixed point retains a \(\frac{1}{2}\ln 2\) entanglement entropy, as shown in Fig.~\ref{fig:gb}(a) of the manuscript.

	\section{Derivation of the \(\mathcal{T}\)-matrix}
	The effective Hamiltonian Eq. (\ref{eq:a1:heff}) consists of the Kondo term, the charge term, the Majorana term, the kinetic term and the Zeeman term,
	\begin{align}
		H_K &= J_K\sum_{\bf kk'} \left[ L^z (c_{{\bf k}\uparrow}^\dagger c_{{\bf k'} \uparrow} - c_{{\bf k}\downarrow}^\dagger c_{{\bf k'}\downarrow} ) + L_+ c_{{\bf k}\downarrow}^\dagger c_{{\bf k'}\uparrow} + L_- c_{{\bf k}\uparrow}^\dagger c_{{\bf k'}\downarrow}  \right], \\
		H_Q &= J_Q \sum_{\bf kk'} \left[ L^z (c_{{\bf k}\uparrow}^\dagger c_{{\bf k'} \uparrow} + c_{{\bf k}\downarrow}^\dagger c_{{\bf k'}\downarrow} ) +L_+ c_{{\bf k}\uparrow}^\dagger c_{{\bf k'}\downarrow}^\dagger + L_- c_{{\bf k}\downarrow} c_{{\bf k'}\uparrow}  \right], \\
		H_M &= J_M \sum_{\bf k} \left[ L^z \gamma (c_{{\bf k}\uparrow} - c_{{\bf k}\uparrow}^\dagger) + L_+\gamma (-c_{{\bf k}\downarrow}^\dagger) + L_- \gamma c_{{\bf k}\downarrow} \right].
	\end{align}
	They can be interpreted in a bilinear way by defining three operators
	\begin{gather}
		K_{\sigma} =  J_K (\sigma L^z c_{0\sigma} + L_{\bar{\sigma}} c_{0\bar{\sigma}}), \\
		Q_{\sigma} = J_Q (c_{0\sigma}L^z + \sigma c_{0\bar{\sigma}}^\dagger L_+), \\
		M_{\uparrow} = M_{\uparrow}^\dagger  = J_M L^z \gamma,\quad M_{\downarrow} = J_M L_+ \gamma,
	\end{gather}
	where \(c_{0\sigma}^{(\dagger)}=\sum_{\bf k}c_{{\bf k}\sigma}^{(\dagger)}\) is the electron annihilation (creation) operator at the origin point. 
	
	The interacting part of the Hamiltonian now can be represented in a compact form 
	\begin{equation}
		H_{\rm int} = \sum_{\sigma} O_{\sigma}^\dagger c_{0\sigma} = \sum_{\sigma} c_{0\sigma}^\dagger O_{\sigma},
	\end{equation}
	where \(O_{\sigma}=K_{\sigma}+Q_{\sigma}+M_{\sigma}\), and one can verify the commutation relations obey
	\begin{equation}
		\begin{split}
			[H,\ c_{{\bf k}\sigma} ] &= -O_{\sigma} -\epsilon_{\bf k}c_{{\bf k}\sigma}, \\
			[H,\ c_{{\bf k}\sigma}^\dagger ] &= O_{\sigma}^\dagger +\epsilon_{\bf k}c_{{\bf k}\sigma}^\dagger.
		\end{split}
	\end{equation}
	The \(\mathcal{T}\)-matrix outputs
	\begin{equation}
		\mathcal{T}_{{\bf kk'}\sigma} = V_{\bf k}^*\begin{bmatrix}
			\braket{\braket{d|d^\dagger}} & -\braket{\braket{d|d}} \\
			-\braket{\braket{d^\dagger|d^\dagger}} & \braket{\braket{d^\dagger|d}} 
		\end{bmatrix} V_{\bf k'}
		= \begin{bmatrix}
			\braket{\braket{O_{\sigma}|O_{\sigma}^\dagger}} & -\braket{\braket{O_{\sigma}|O_{\sigma}}} \\
			-\braket{\braket{O_{\sigma}^\dagger|O_{\sigma}^\dagger}} & \braket{\braket{O_{\sigma}^\dagger|O_{\sigma}}} 
		\end{bmatrix}
		+ (\sigma J_K + J_Q) \braket{L^z}.
	\end{equation}
	
	\section{Finite-temperature correlation functions within FDM-NRG}
	In this section we summarize the full-density-matrix (FDM) formulation used to evaluate finite-temperature correlation functions along the Wilson chain. The derivation follows the complete-basis construction of discarded states in NRG~\cite{weichselbaum2012tensor,merker2012fdm}.
	
	\subsection{Complete basis and FDM}
	The discarded states of all NRG iterations form a complete orthonormal basis. The completeness and orthogonality relations can be written as
	\begin{gather}
		\mathbbm{1} = \sum_{n=n_0}^{N}\sum_{l,e} \ket{n;l,e}\bra{n;l,e}, \\
		\braket{n;l,e|n';l',e'} = \delta_{nn'}\delta_{ll'}\delta_{ee'} ,
	\end{gather}
	where \(l\) labels the discarded eigenstates of the iteration-\(n\) Hamiltonian \(H^{(n)}\), \(e\) labels the environment degrees of freedom, and \(n_0\) is the first iteration where truncation is performed. The factorized form reads \(\ket{n;l,e}=\ket{l}_n\otimes\ket{e}_n\). The eigenstates at the last iteration \(N\) are also treated as discarded, so that the above sum defines a complete basis.
	
	On this basis, the full density matrix is defined by
	\begin{equation}
		\rho = \sum_{n=n_0}^{N}\sum_{l,e} \ket{n;l,e}\frac{e^{-\beta E_{n;l}}}{Z}\bra{n;l,e},
		\label{eq:app:fdm_rho}
	\end{equation}
	where \(E_{n;l}\) is the discarded-state energy at iteration \(n\), and the full partition function is
	\begin{equation}
		Z(T)=\sum_{n=n_0}^{N} d^{N-n}\sum_{l} e^{-\beta E_{n;l}} .
		\label{eq:app:fdm_Z}
	\end{equation}
	Here \(d\) is the local dimension of a Wilson-chain site (\(d=4\) for a single-channel spinful fermionic bath). It is convenient to introduce a normalized shell density matrix
	\begin{equation}
		\tilde{\rho}_n = \sum_{l,e} \ket{n;l,e}\frac{e^{-\beta E_{n;l}}}{\tilde{Z}_n}\bra{n;l,e},
		\qquad
		\tilde{Z}_n(T)=d^{N-n} Z_n(T),\ \ Z_n(T)=\sum_{l} e^{-\beta E_{n;l}} ,
		\label{eq:app:fdm_rho_tilde}
	\end{equation}
	so that the FDM takes a weighted sum form,
	\begin{equation}
		\rho=\sum_{n=n_0}^{N} w_n \tilde{\rho}_n,
		\qquad
		w_n=d^{N-n}\frac{Z_n}{Z},\ \ \sum_{n=n_0}^{N} w_n=1 .
		\label{eq:app:fdm_weights}
	\end{equation}
	
	\subsection{Thermal averages and two-point correlators}
	For an operator \(\hat{A}\) acting only within the subsystem diagonalized at iteration \(n\) (i.e., independent of the environment), the thermal average can be written as
	\begin{equation}
		\braket{\hat{A}}
		=\mathrm{Tr}(\rho\hat{A})
		=\sum_{n=n_0}^{N}\sum_{l} w_n A_{ll}^{(n)} \frac{e^{-\beta E_{n;l}}}{Z_n},
		\label{eq:app:fdm_A}
	\end{equation}
	where \(A_{ll}^{(n)}=\braket{n;l|\hat{A}|n;l}\) and we have used the fact that \(\hat{A}\) does not act on the environment degrees of freedom. The same logic extends to multi-point operators.
	
	As an example, consider a two-point correlator between an impurity operator \(\hat{S}^z_0\) and a bath-site operator \(\hat{S}^z_k\) on the Wilson chain. One obtains
	\begin{equation}
		\braket{\hat{S}^z_0\hat{S}^z_k}
		=\sum_{n=n_0}^{N}\sum_{l} w_n \left(S^z_0S^z_k\right)_{ll}^{(n)} \frac{e^{-\beta E_{n;l}}}{Z_n}.
		\label{eq:app:fdm_SzSz}
	\end{equation}
	The matrix element \(\left(S^z_0S^z_k\right)_{ll}^{(n)}\) vanishes if the second operator is located outside the subsystem diagonalized at iteration \(n\), i.e., \(k>n\), because \(\hat{S}^z_k\) then acts only on the environment part of \(\ket{n;l,e}\). Concretely,
	\begin{equation}
		\hat{S}^z_k\ket{n;l,e}
		=\ket{l}_n\otimes \left(S^z(e)\ket{e}_n\right),
	\end{equation}
	where \(S^z(e)\) is the \(U(1)\) spin quantum number on site \(k\) in the environment basis \(\{\ket{0},\ket{\downarrow},\ket{\uparrow},\ket{\uparrow\downarrow}\}\).
	
	Finally, the spatially integrated correlator (used in the main text, with the sum running over bath sites) reads
	\begin{equation}
		\sum_{k=1}^{N}\braket{\hat{S}^z_0\hat{S}^z_k}
		=\sum_{k=1}^{N}\sum_{n=\max\{k,n_0\}}^{N}\sum_{l} w_n \left(S^z_0S^z_k\right)_{ll}^{(n)} \frac{e^{-\beta E_{n;l}}}{Z_n}.
		\label{eq:app:fdm_integrated_corr}
	\end{equation}
	This expression applies equally to vector correlators such as \(\sum_k \langle \mathbf{L}_{\rm QD}\cdot \mathbf{S}_k\rangle\) and \(\sum_k \langle \mathbf{L}_{\rm QD}\cdot \bm{\eta}_k\rangle\) by replacing \(\hat{S}^z\) with the appropriate local operators and evaluating the corresponding diagonal matrix elements at each iteration.

	\end{widetext}

\end{document}